\documentclass[aps,preprint,amsfonts,superscriptaddress,byrevtex,nofootinbib,showpacs]{revtex4}
\newcommand{\beq}{\begin{equation}}
\newcommand{\eeq}{\end{equation}}

\def\bm#1{\mbox{\boldmath{$#1$}}}
\usepackage{graphicx}
\usepackage{feynmf}
\usepackage{bm}
\begin{document}
\title{Gauge fixing, BRS invariance
and Ward identities for randomly stirred flows}
\author{Arjun Berera}
\email{ab@ph.ed.ac.uk}  \affiliation{School of Physics and
Astronomy, University of Edinburgh, Edinburgh, EH9 3JZ, U.K}
\author{David Hochberg}
\email{hochbergd@inta.es} \affiliation{Centro de Astrobiolog\'{\i}a
(CSIC-INTA), Ctra. Ajalvir Km. 4, 28850 Torrej\'{o}n de Ardoz,
Madrid, Spain}

\date{\today}
\begin{abstract}
The Galilean invariance of the Navier-Stokes equation is shown to be
akin to a global gauge symmetry familiar from quantum field theory.
This symmetry leads to a multiple counting of infinitely many
inertial reference frames in the path integral approach to randomly
stirred fluids. This problem is solved by fixing the gauge, i.e.,
singling out one reference frame. The gauge fixed theory has an
underlying Becchi-Rouet-Stora (BRS) symmetry which leads to the Ward
identity relating the exact inverse response and vertex functions.
This identification of Galilean invariance as a gauge symmetry is
explored in detail, for different gauge choices and by performing a
rigorous examination of a discretized version of the theory. The
Navier-Stokes equation is also invariant under arbitrary rectilinear
frame accelerations, known as extended Galilean invariance (EGI). We
gauge fix this extended symmetry and derive the generalized Ward
identity that follows from the BRS invariance of the gauge-fixed
theory. This new Ward identity reduces to the standard one in the
limit of zero acceleration. This gauge-fixing approach unambiguously
shows that Galilean invariance and EGI constrain only the
\textit{zero mode} of the vertex but none of the higher wavenumber
modes.
\end{abstract}

\pacs{47.27.ef, 11.10.-z, 03.50.-z}
\date{\today}

\maketitle

\section{\label{sec:intro}Introduction}

The formulation of the Navier Stokes equation with random forcing as
a classical stochastic field theory
\cite{DeDom,Mou,Teo,Toma,DM,Eyink,BH} opens up the way to apply the
methods originally developed for quantum fields \cite{ZJ}. This fact
has been used in the majority of cases for applying renormalization
group (RG) methods to models of fully developed turbulence
\cite{AAV}. This field theory approach has also been used on many
occasions to derive exact relations between different correlation
functions implied by Galilean invariance. These are akin to the
Ward-Takahashi identities (WTI) of quantum field theory
\cite{Ramond}. The most well known of these exact identities relates
the vertex and response functions \cite{DM}.  Based on this
relation, claims have been made concerning the non-renormalization,
under the renormalization group, of the advective or inertial term
in the Navier-Stokes equation \cite{FNS,Teo,Toma,Mou,DM,Mc,BH,BH2}.
Despite the long history of applying these field theoretic methods,
two major problems connected with the Galilean invariance of the
path integral framework of randomly stirred fluids have been brought
to light recently. The first problem has to do with the functional
itself. The standard dynamic functional for the randomly forced
Navier-Stokes equation leads to spurious relations for the
correlation functions involving the zero-mode of the fluid velocity.
This dynamic functional is therefore ill-defined. As shown in
\cite{BH2} the dynamic functional for the randomly stirred Navier
Stokes equation can be regarded as a gauge theory, and then the
problem of the spurious relations is solved by fixing the gauge,
which thus breaks the Galilean invariance. This is a new and
important observation and leads to a well-defined functional
expression, which has been missing for the past three decades.
However, once the gauge is fixed, it is crucial to verify whether
Galilean invariance (GI) can be restored, and this is achieved by
recognizing that the gauge-fixed theory is BRS invariant. This
brings us to the second problem which has to do with the past
mis-interpretations of the Ward identity that follows from the
functional. This BRS invariance leads to a Ward identity whose
physical consequence is that \textit{only} the zero mode part of the
full vertex, but none of the higher spatial modes, is constrained.
In other words, contributions from the non-zero modes to the vertex
can receive corrections under renormalization. This conclusion
therefore provides greater clarity to, and in some cases goes
against, previous interpretations regarding the physical
consequences of GI for the vertex in the Navier-Stokes equation
\cite{FNS,Teo,Toma,Mou,DM,Mc,BH,BH2}.

The present work supports the conclusions drawn in \cite{BH} in
\cite{Mc} and in \cite{BH2}, and develops in full detail the results
presented in \cite{BH2}, while also generalizing that work. For
this, the gauge fixing concept for the NSE is explored in detail
including the treatment of distinct gauge choices and the associated
implications for the BRS symmetry left in the theory. Also, the
analogy between the Ward identities found here for the NSE and those
in QED is carefully examined. Finally, the considerations developed
so far for Galilean invariance are thoroughly worked out here for
extended Galilean invariance, which is a more general global
symmetry of the NSE. We derive the consequences of this higher
symmetry which leads to a new Ward identity for the forced NSE. The
Ward identity for EGI has not to our knowledge been derived nor
applied to the Navier-Stokes equation before. We then use it to see
how the convective term in the Navier Stokes equation can
renormalize. This has direct physical consequences. For this case,
the time dependent but spatial zero mode part of the vertex is
related to viscosity and mass renormalizations, while none of the
higher spatial modes are constrained by EGI. These facts constitute
the physical results of this paper.

In order to consider the consequences of breaking Galilean
invariance via gauge fixing, we start with the ensemble of histories
of Navier-Stokes dynamics, and make direct use of the standard
functional integral methods. The dynamic generating functional for a
randomly stirred incompressible fluid has been extensively studied
for many years \cite{DeDom,Mou,Teo,Toma,DM,Eyink,BH}. It is based on
the path integral approach \cite{DeDom,Jan,Phythian,Jensen} to
classical statistical dynamics \cite{MSR} and is given by
\begin{equation}\label{Z}
Z = \int [D\mathbf{V}][D\bm{\sigma}] \exp \{ -S[\mathbf{V},
\bm{\sigma}] \},
\end{equation}
where the action is \footnote{A Martin-Siggia-Rose type action can
be derived for general Langevin equations of motion
\cite{DeDomPeliti}, and this includes the Navier-Stokes equation
(NSE) with random forcing. The Jacobian that arises in passing from
the general stochastic equation to the path integral can, in effect,
\textit{always} be set to unity. This is treated in detail in Ref.
\cite{DeDomPeliti}. See also Ref. \cite{Jensen} and Refs. \cite{Mou,
Eyink} which prove that this Jacobian is unity for the particular
case of the forced NSE.}
\begin{eqnarray}\label{class}
S[{\bf V}, \bm{\sigma}] &=& \frac{1}{2}\int d{\bf x} dt \int d{\bf
y}\, {\bf \sigma}_i({\bf x},t)D_{ij}({\bf x} - {\bf y}) {\bf
\sigma}_j({\bf y},t)\nonumber \\
 && -i \int d{\bf x} dt\,\, \sigma_k
\Big(\frac{\partial V_k}{\partial t} + P_{kj}(\nabla )
\frac{\partial (V_lV_j)}{\partial x_l} - \nu_0 \nabla^2 V_k\Big).
\end{eqnarray}
The instantaneous fluid velocity is $\mathbf{V}$, the conjugate
field is $\bm{\sigma}$;  $D_{ij}({\bf x})$ is the only non-vanishing
cumulant of the stationary stirring force and $\mathbf{x}$ is a
coordinate vector in $d$-dimensions. The projection operator is
$P_{ij}(\nabla ) = \big(\delta_{ij} - \nabla_i \frac{1}{\nabla^2}
\nabla_j \big)$ and $\nu_0$ denotes the bare (unrenormalized) fluid
viscosity. Now consider a second primed frame moving with a constant
velocity $\textbf{c}$ with respect to the unprimed frame. Then the
relations between the instantaneous fluid velocity, conjugate field,
time and coordinates of events in both frames are given by:
\begin{eqnarray}
\label{galZ11} \mathbf{V^{c}}({\bf x}, t) &=& {\mathbf V'}({\bf
x'},t') + \mathbf{c},
\\
\label{galZ21} \bm{\sigma^{c}}({\bf x}, t) &=& \bm{\sigma}'({\bf
x'},t'),\\ \label{galZ31}
t &=& t', \\
\label{galZ41} {\bf x} &=& {\bf x}' +{\bf c}t.
\end{eqnarray}
$\mathbf{V^{c}}$ denotes the result of the Galilean transformation
applied to the velocity field $\mathbf{V}$, and similarly for
$\bm{\sigma^{c}}$. Due to the Galilean invariance of the measure
$[D\mathbf{V^c}][D\bm{\sigma^c}] = [D\mathbf{V'}][D\bm{\sigma'}]$
\footnote{The change of variables Eq.(\ref{galZ11}) in the
functional measure generates a unit jacobian: ${\cal J} = \det
\frac{\delta V_i(\mathbf{x},t)}{\delta V_j'(\mathbf{y'},\tau')} =
\det \frac{\delta (V'_i(\mathbf{x'},t')+ c_i)}{\delta
V_j'(\mathbf{y'},\tau')} = \det (
\delta_{ij}\delta(\mathbf{x'}-\mathbf{y'})\delta(\tau'-t') ).$
Similarly for the transformation Eq.(\ref{galZ21}) of the conjugate
field. The Jacobian is also unity for the extended Galilean
transformation in Eqs.(\ref{exgaltrans1}-\ref{exgaltrans4}) and
Eq.(\ref{exgaltrans5}). } and the action $S$ under the
transformation in Eqs.(\ref{galZ11}-\ref{galZ41}), it is clear that
the generating functional Eq.(\ref{Z}) sums over all fluid velocity
configurations. This results in multiple counting of
\textit{physically equivalent configurations}: namely, those that
are equivalent up to a Galilean transformation (GT). In other words,
this integral includes the sum over all inertial reference frames.
This feature leads to spurious relations among velocity correlation
functions which must be removed in order to render a well defined
generating functional for the theory \cite{BH2}. The solution to
this problem was demonstrated to be provided by gauge fixing
\cite{BH2}. Moreover, if one wants to use the path integral for the
Navier-Stokes equation for computing non-Galilean invariant
quantities, like $n$-point velocity correlation  functions, it is
essential to carry out a gauge fixing along the lines we present
here, and this is especially true for numerical implementations
\cite{Doben}.  Gauge fixing for $Z$ can be viewed operationally in
exactly the same way as in quantum field theory. In the present
context, we must divide velocity configuration space into
equivalence classes called the ``orbits" of the GT.  An orbit of the
GT includes all velocity field configurations which result when all
possible GT's are applied to a given initial velocity field
configuration. Thus, $Z$ is proportional to the infinite volume of
these orbits, and this volume factor is extracted out before
defining this functional. However, once the Galilean invariance is
explicitly broken, it is crucially important to verify whether the
exact identities relating inverse response and vertex functions
continue to hold. We demonstrate that this is the case, because the
resultant gauge fixed functional has an underlying
Becchi-Rouet-Stora (BRS) symmetry. This symmetry leads to exact
identities structurally similar to the Slavnov-Taylor (ST)
identities of quantum field theory \cite{BRS}. Here, the
Slavnov-Taylor identity is significant because the ghost degrees of
freedom make absolutely explicit that the vertex nonrenormalization
only affects the zero mode. Contrary to previous assertions and
claims, it is not even in the $k \rightarrow 0$ limit that the
claimed vertex nonrenormalization holds, but rather \textit{only} at
$k=0$, the zero-mode. This is a subtle but very important point that
must be born in mind.

The outline of this paper is as follows. In Sec.
\ref{sec:FPdeterminant} we pick a gauge which singles out a unique
inertial reference frame, and carry out the gauge fixing procedure
for the dynamic functional in Eq. (\ref{Z}). This leads to a gauge
fixed action. In Sec. \ref{sec:BRSsym} we write down the
Becchi-Rouet-Stora (BRS) transformation that leaves this gauge fixed
action invariant. This BRS invariance leads to the Slavnov-Taylor
identities for the dynamic functional. These in turn imply the
crucial Ward identity for the effective action that has been derived
previously on numerous occasions, but always derived for the
\textit{non-gauge} fixed functional \cite{FNS,Teo,Toma,Mou,DM,BH}.
In Sec \ref{sec:ST} we prove that this crucial identity remains
valid, even after explicitly breaking the Galilean invariance, and
selecting a fixed reference frame. Other gauge choices are possible,
as illustrated in Sec \ref{sec:deltagf}. The stochastic field theory
for the NSE can be regularized in a space-time box and a
corresponding Ward identity is derived in Sec \ref{sec:boxward}.
This is used to demonstrate that the spurious relations first found
in \cite{BH2} for the continuum field theory arise in the
regularized theory as well, and are therefore not an artifact of the
continuum limit. The Navier-Stokes equation is also invariant under
\textit{arbitrary} rectilinear frame accelerations, a symmetry known
as extended Galilean invariance (EGI) \cite{Pope}. This is reviewed
briefly in Sec \ref{sec:egiNSE}. Gauge fixing is carried out in Sec
\ref{sec:FPdetEGI} by choosing a single accelerating frame and the
BRS transformation that leaves the corresponding gauge fixed action
invariant is written down in Sec \ref{sec:BRSEGI}. To complete the
analysis, we derive the Slavnov-Taylor identities for EGI in Sec
\ref{sec:WTIEGI} and show that these lead to a new identity for the
inverse response and vertex functions. This identity reduces to the
standard one in the limit of vanishing reference frame
accelerations. The Ward identities that follow from the EGI and GI
in stochastic flows bear a striking resemblance to the well known
QED Ward identities that follow from the local U(1) gauge
invariance. The similarities as well as the differences are spelled
out in Sec \ref{sec:QED}. We discuss the implications of both
Galilean and extended Galilean invariance for vertex renormalization
in Sec \ref{sec:vertex} and present our conclusions in Sec
\ref{sec:disc}. The main steps for gauge fixing the Galilean
invariant dynamic functional are collected in an Appendix which
closely parallel the standard technique as employed in quantum field
theory.

\section{\label{sec:gi} Galilean Invariance: Gauge fixing and Becchi-Rouet-Stora Symmetry}

\subsection{\label{sec:FPdeterminant} The $\mathbf{V_0}^2$ gauge}

For the reasons stated above, we proceed to gauge-fix the dynamic
functional in Eq.(\ref{Z}). The basic outline for doing so in an
arbitrary gauge $\mathbf{f}(\mathbf{V})$ is summarized in general
terms in the Appendix \ref{appendix}. This is intended to parallel
closely the procedure as it is used in quantum gauge theories. Once
we choose the specific gauge introduced below, we then only have to
calculate the corresponding Fadeev-Popov determinant
Eq.(\ref{FPdet}), and then select the arbitrary function
$\mathbf{U}$ that appears in Eq.(\ref{pathint2}) to complete the
procedure. Here and below, we consider spatially bounded fluid
systems.

Recall that the purpose of gauge fixing here is to single out one
inertial reference frame. This can be accomplished by constraining
the \textit{zero mode} part of the full instantaneous velocity
field, because this mode corresponds to the constant velocity of the
system as a whole, that is, its bulk velocity.  We thus make the
gauge choice specified by
\begin{equation}\label{gaugechoice}
\mathbf{f}(\mathbf{V}) = \frac{1}{vol}\int d\mathbf{x}dt \,
\mathbf{V}(\mathbf{x},t),
\end{equation}
where $vol = L^3T$ is the (finite) volume of the space-time box
bounding the fluid. This projects out the zero mode from the full
instantaneous velocity field $\mathbf{V}$, as required.

Next, from  Eqs.(\ref{galZ11},\ref{galZ31},\ref{galZ41}), we have
\begin{eqnarray}
V_j^{\mathbf{c}}(\mathbf{x},t) &=& V'_j(\mathbf{x}-\mathbf{c}t,t) +
c_j, \nonumber \\
&=& V'_j(\mathbf{x},t) -c_kt\frac{\partial
V'_j(\mathbf{x},t)}{\partial x_k} + c_j + O(c^2),\nonumber \\
\delta V_j^{\mathbf{c}}(\mathbf{x},t) &=& V'_j - V^{\mathbf{c}}_j =
c_kt\frac{\partial V'_j(\mathbf{x},t)}{\partial x_k} - c_j,
\nonumber
\\ \label{derivative} {\rm Thus} \qquad \frac{\partial
V_j^{\mathbf{c}}(\mathbf{x},t)}{\partial c_m}\lfloor_{\mathbf{c} =
\mathbf{0}} &=&  t\frac{\partial V'_j(\mathbf{x},t)}{\partial x_m} -
\delta_{jm}.
\end{eqnarray}
Then from Eqs.(\ref{gaugechoice},\ref{derivative}) we have
\begin{equation}
\Big(\frac{\partial f_i}{\partial c_j}\Big)|_{\mathbf{c} =
\mathbf{0}} = \frac{1}{vol}\int d\mathbf{x}dt \, \Big(
t\frac{\partial V'_i(\mathbf{x},t)}{\partial x_j} -
\delta_{ij}\Big).
\end{equation}
This simplifies due to boundary conditions. Imposing the physically
reasonable periodic boundary conditions for a spatial box of side
length $L$, we have
\begin{equation}
\int_{0}^L dx_j \frac{\partial V_i(\mathbf{x},t)}{\partial x_j} =
V_i(\mathbf{x}_ {\perp},x_j=L,t) - V_i(\mathbf{x}_{\perp},x_j=0,t) =
0,
\end{equation}
where $\mathbf{x}_ {\perp}$ are the coordinates orthogonal to $x_j$.
Thus for the Fadeev-Popov determinant in Eq.(\ref{FPdet}), we
calculate
\begin{equation}\label{FP2}
\Delta_f[\mathbf{V}] = \det \Big(\frac{\partial f_i}{\partial
c_j}\Big)|_{\mathbf{c} = \mathbf{0}} = \det(-\delta _{ij}) = \int
d\bm{\eta} d\bm{\eta}^* \exp\Big(-i \bm{\eta}^* \cdot \bm{\eta}
\Big).
\end{equation}
The last equality expresses this determinant as an integral over
constant complex conjugate Grassmann vectors $\bm{\eta}$ and
$\bm{\eta}^*$, with $\{ \eta_i,\eta^*_j\} = 0$ \cite{Ramond}.
Lastly, since our gauge $\mathbf{f}(\mathbf{V})$
Eq.(\ref{gaugechoice}) is simply a \textit{constant vector}, we
repeat the arguments leading to Eq.(\ref{pathint2}) and
Eq.(\ref{gfixed}) in the Appendix, but here replace the function
$\mathbf{U} (\mathbf{x},t) \rightarrow \mathbf{b}$ by a constant
vector in Eq.(\ref{popchoice}), and drop the spacetime integrations
indicated there, that is, instead of Eq.(\ref{popchoice}) we choose
$G[\mathbf{b}] = \exp\Big( -\frac{1}{2\xi} \mathbf{b}^2 \Big)$,
where $\xi > 0$ is a free parameter.

Putting these results together, and after integrating over the delta
-function in Eq.(\ref{pathint2}), which in this case is
$\delta(\mathbf{V_0}-\mathbf{b})$, the gauge fixed action reads:
\begin{equation}\label{GFactionxt}
S_{GF} = S + \frac{1}{2\xi}\mathbf{V_0}^2
+i\bm{\eta}^*\cdot\bm{\eta},
\end{equation}
where $S$ is given in Eq.(\ref{class}) and we \textit{define}
$\mathbf{V_0} \equiv \frac{1}{vol}\int d\mathbf{x}dt \,
\mathbf{V}(\mathbf{x},t)$, and note that this has the correct
dimensions of a velocity. Galilean invariance is manifestly broken
in $S_{GF}$.

\subsection{\label{sec:BRSsym} Becchi-Rouet-Stora invariance of the gauge-fixed action}

Despite the fact that Galilean invariance has been broken
explicitly, the gauge-fixed theory Eq.(\ref{GFactionxt}) does have a
fundamental underlying symmetry which makes full use of the
Grassmann vectors introduced to represent the Fadeev-Popov
determinant Eq.(\ref{FP2}). This symmetry was originally established
to be a \textit{general consequence} of gauge fixing in non-Abelian
gauge theories, and is known as Becchi-Rouet-Stora (BRS)
invariance\footnote{The BRS symmetry in our paper arises from
gauge-fixing, and should thus be distinguished clearly from the BRS
symmetry that arises from regarding the stochastic differential
equation itself as the constraint; see, e.g., Ref. \cite{ZJ}.}
\cite{BRS}.

Consider the following infinitesimal BRS transformation where
$\zeta^2 = 0$ is a real Grassmann constant:
\begin{eqnarray}\label{infBRSx}
\delta_{\rm BRS} \mathbf{x} &=& -\zeta(\bm{\eta} + \bm{\eta}^*),
\\\label{infBRSt} \delta_{\rm BRS} t &=& 0,
\\\label{infBRSV} \delta_{\rm BRS} \mathbf{V}(\mathbf{x},t) &=&
\zeta(\eta_k + \eta_k^*) t\frac{\partial
\mathbf{V}(\mathbf{x},t)}{\partial x_k} - \zeta(\bm{\eta} +
\bm{\eta}^*), \\\label{infBRSsig} \delta_{\rm
BRS}\bm{\sigma}(\mathbf{x},t) &=& \zeta(\eta_k + \eta_k^*)
t\frac{\partial \bm{\sigma}(\mathbf{x},t)}{\partial x_k},
\\\label{infBRSeta}
\delta_{\rm BRS} \bm{\eta} &=& -\frac{i}{\xi}\mathbf{V_0}\zeta, \\
\label{infBRSeta*} \delta_{\rm BRS} \bm{\eta}^* &=&
+\frac{i}{\xi}\mathbf{V_0}\zeta.
\end{eqnarray}
Note that Eq.(\ref{infBRSV}) automatically implies the corresponding
BRS transformation for the velocity zero mode $\mathbf{V_0}$:
\begin{equation}\label{infBRSzero}
\delta_{BRS}\mathbf{V_0} = \frac{1}{vol}\int d\mathbf{x}dt \,
\delta_{BRS}\mathbf{V}(\mathbf{x},t) = -\zeta(\bm{\eta} +
\bm{\eta}^*).
\end{equation}
Now Eqs.(\ref{infBRSx}-\ref{infBRSsig}) are recognized to be just
the \textit{infinitesimal} version of the Galilean transformation
Eqs.(\ref{galZ11}-\ref{galZ41}), but with the boost velocity
replaced by the Grassmann vectors $\mathbf{c} \rightarrow
\zeta(\bm{\eta} + \bm{\eta}^*)$. Therefore, the action $S$ and path
integral measure are automatically invariant under this subset of
the full BRS transformation. Next, from
Eqs.(\ref{infBRSeta},\ref{infBRSeta*},\ref{infBRSzero}), we easily
verify that the combined gauge fixing and Grassmann terms in
$S_{GF}$ are invariant. From these observations we therefore
conclude that the complete gauge fixed action $S_{GF}$ in
Eq.(\ref{GFactionxt}) is invariant under the full BRS transformation
Eqs.(\ref{infBRSx}-\ref{infBRSeta*}):
\begin{equation}
\delta_{BRS}S_{GF} = 0 .
\end{equation}
This BRS invariance leads to exact identities which we derive below.

\subsection{\label{sec:ST} The Slavnov-Taylor and Ward identities}

We introduce source terms into the gauge-fixed functional as
follows:
\begin{eqnarray}\label{gfZxt}
Z_{GF}[\mathbf{J},\bm{\Sigma},\bm{\theta},\bm{\theta}^*] &=& \int
[D{\bf V}][D \bm{\sigma}]d\bm{\eta}d\bm{\eta}^*\, \exp\{-S_{GF}[{\bf
V}, \bm{\sigma},
\bm{\eta},\bm{\eta}^*] + \bm{\theta}^*\cdot\bm{\eta} + \bm{\theta}\cdot\bm{\eta}^* \nonumber \\
&+& \int d\mathbf{x} dt \,
\{\mathbf{J}(\mathbf{x},t)\cdot\mathbf{V}(\mathbf{x},t) +
\mathbf{\Sigma}(\mathbf{x},t)\cdot\bm{\sigma}(\mathbf{x},t) \} \},
\end{eqnarray}
where $\bm{\theta}$ and $\bm{\theta}^*$ are complex Grassmann
vectors. Next, we subject this functional to the complete BRS
transformation (that is, we displace all fields by the infinitesimal
BRS transformation) Eqs.(\ref{infBRSx}-\ref{infBRSeta*}). Since the
measure and the gauge-fixed action $S_{GF}$ are BRS-invariant, only
the source terms in Eq.(\ref{gfZxt}) will be affected. Moreover, as
$\zeta^2 = 0$, we can easily expand the exponential: $\exp(\zeta A)
= 1 + \zeta A$. The gauge-fixed functional Eq.(\ref{gfZxt}) thus
transforms as $Z_{GF} \rightarrow Z_{GF} + \delta_{BRS}Z_{GF}$,
where $\delta_{BRS}Z_{GF} = 0$ can be written as follows:
\begin{eqnarray}\label{deltaBRS}
\left[ \frac{i}{\xi}(\bm{\theta}^* - \bm{\theta})\cdot
\frac{1}{vol}\int d\mathbf{x}dt \, \frac{\delta}{\delta
\mathbf{J}(\mathbf{x},t)} + \big(\frac{\partial}{\partial \theta_j}
+\frac{\partial}{\partial \theta_j^*}\big){\hat O}_{j} \right]
Z_{GF}[\mathbf{J},\bm{\Sigma},\bm{\theta},\bm{\theta}^*] = 0.
\label{WTIgf}
\end{eqnarray}
Here the operator
\begin{eqnarray}
{\hat O}_{j}&=& \int d\mathbf{x} dt \, \Big( J_m t\nabla_j
\frac{\delta}{\delta J_m} + \Sigma_m t\nabla_j \frac{\delta}{\delta
\Sigma_m} - J_j \Big) \label{OWT}
\end{eqnarray}
is the same operator appearing in the well known Ward identity that
follows from the Galilean invariance of the \textit{non-gauge} fixed
functional \cite{DM,Teo,Mou,Toma}: that is, ${\hat
O}_{j}Z[\bm{J},\bm{\Sigma}] = 0$.

To finish, we introduce the effective action $\Gamma$ and write the
Slavnov-Taylor identity Eq.(\ref{deltaBRS}) in terms of this
quantity. Let $W = \ln Z_{GF}$, then the generating functional of
one-particle irreducible functions is given by the Legendre
transform
\begin{equation}\label{gamma0}
\Gamma[\mathbf{V}_{cl},\bm{\sigma}_{cl},\bm{\eta}_{cl},\bm{\eta}_{cl}^*]
= -W[\mathbf{J}, \bm{\Sigma},\bm{\theta},\bm{\theta}^*] +
\bm{\theta^*\cdot\eta}_{cl} + \bm{\theta\cdot\eta}^*_{cl} + \int
d\mathbf{k}d\omega \, (\mathbf{J}\cdot\mathbf{V}_{cl} +
\mathbf{\Sigma}\cdot\bm{\sigma}_{cl}),
\end{equation}
which for convenience, we express in wavevector $\mathbf{k}$ and
frequency space $\omega$. Here, the label ``$cl$" reminds us that
the fields so indicated are averaged over the fluctuating fields and
in the presence of the source terms: that is, using the ensemble in
Eq.(\ref{gfZxt}). From Eq.(\ref{gamma0}) we have that
\begin{equation}
J_k = \frac{\delta \Gamma}{\delta V_k^{cl}},\,\, \Sigma_k =
\frac{\delta \Gamma}{\delta \sigma_k^{cl}},\,\, \theta_j =
\frac{\partial \Gamma}{\partial {\eta^*}^{cl}_j},\,\, \theta^*_j =
\frac{\partial \Gamma}{\partial {\eta}^{cl}_j}.
\end{equation}
We can now straightforwardly write down the identity for $\Gamma$
that follows directly from $\delta_{BRS}Z_{GF} = 0$:
\begin{eqnarray}\label{WTI2}
&& 0 = \frac{i}{\xi}\mathbf{V}^{cl}_0\cdot \Big( \frac{\partial
\Gamma}{\partial \bm{\eta}_{cl}} - \frac{\partial
\Gamma}{\partial \bm{\eta}^*_{cl}} \Big) + (\eta_{cl} + \eta^*_{cl})_j\times \\
&& \int d\mathbf{k} d\omega \, \Big(k_j \frac{\partial
V^{cl}_m(\mathbf{k},\omega)}{\partial \omega} \frac{\delta \Gamma}
{\delta V^{cl}_m(\mathbf{k},\omega)} + k_j \frac{\partial
\sigma^{cl}_m(\mathbf{k},\omega)}{\partial \omega} \frac{\delta
\Gamma} {\delta \sigma^{cl}_m(\mathbf{k},\omega)} -
\delta(\mathbf{k})\delta(\omega)\frac{\delta \Gamma} {\delta
V^{cl}_j(\mathbf{k},\omega)}\Big) \nonumber.
\end{eqnarray}
We now make use of this formula and apply it to the problem at hand.
The dependence of $\Gamma$ on $\bm{\eta}$ and $\bm{\eta}^*$ is
simple since these constant Grassmann vector fields do not interact
nor do they couple to the velocity or conjugate fields \footnote{
This dependence of $\Gamma$ on the Grassmann variables is similar to
that for gauge-fixed quantum electrodynamics, see, e.g. the
development in Ramond's book \cite{Ramond}.}. Thus, we can
immediately write
\begin{equation}\label{Gamma2}
\Gamma[\mathbf{V}_{cl},\bm{\sigma}_{cl},\bm{\eta}_{cl},\bm{\eta}_{cl}^*]
= -i \bm{\eta}_{cl}^* \cdot \bm{\eta}_{cl} +
\Gamma[\mathbf{V}_{cl},\bm{\sigma}_{cl}],
\end{equation}
where $\Gamma[\mathbf{V}_{cl},\mathbf{\sigma}_{cl}]$ does not depend
on either $\bm{\eta}_{cl}$ or $\bm{\eta}_{cl}^*$, and the first few
required terms are written out in Eq.(\ref{GammaTaylor}) in Appendix
\ref{sec:gammaexp}.

The pertinent identity we seek is then obtained by inserting
Eq.(\ref{Gamma2}) into Eq.(\ref{WTI2}), and then differentiating
this with respect to $\delta/\delta
V^{cl}_l(\mathbf{k},\omega)\delta/\delta
\sigma^{cl}_n(-\mathbf{k},-\omega)$, and then setting
$\mathbf{V}^{cl}$ = $\bm{\sigma}^{cl}= 0$. Note that the terms in
Eq.(\ref{WTI2}) depending on the gauge parameter $\xi$ do not
contribute to this sequence of steps, and most importantly, we end
up obtaining the following result:
\begin{equation}\label{WTIk2}
(\eta_{cl} + \eta^*_{cl})_j\Big(k_m\frac{\partial }{\partial
\omega}\Gamma_{ln}^{(1,1)}(\mathbf{k},\omega; -\mathbf{k},-\omega) +
\Gamma_{mln}^{(2,1)} (\mathbf{0},0; \mathbf{k},\omega;
-\mathbf{k},-\omega)\Big) = 0.
\end{equation}
Moreover, since $\bm{\eta}_{cl}$ and $\bm{\eta}^*_{cl}$ are
\textit{arbitrary}, the expression within the larger parentheses
must vanish identically. $\Gamma^{(1,1)}$ and $\Gamma^{(2,1)}$
denote the inverse response and vertex functions, respectively.
Thus, we recover the well-known Ward identity derived previously on
many occasions from the non-gauge fixed action
\cite{Teo,Mou,Toma,BH}. Alternatively, we can differentiate
Eq.(\ref{WTI2}) with respect to $\partial/{\partial \eta^{cl}_j}\,
\delta/\delta V^{cl}_l(\mathbf{k},\omega) \delta/\delta
\sigma^{cl}_n(-\mathbf{k},-\omega)$ followed by setting
$\bm{\eta}^{cl} = \bm{\eta}^*_{cl} = \mathbf{V}^{cl}$ =
$\bm{\sigma}^{cl}= \mathbf{0}$. The end result is the same.

In summary, the Ward identity Eq.(\ref{WTIk2}) relating the exact
inverse response function to the exact vertex of the NSE holds for
the gauge fixed action Eq.(\ref{GFactionxt}), is a direct
consequence of the BRS invariance of the latter, and is
\textit{independent} of the gauge parameter $\xi$. From this point
on, the entire discussion regarding the implications of Galilean
invariance for the vertex renormalization \cite{BH} continues to
apply; see Sec \ref{sec:vertex}. Most importantly, we see that those
arguments are valid \textit{even after breaking the Galilean
invariance and going to a fixed reference frame}.

\subsection{\label{sec:deltagf} The $\delta$-gauge}

Other gauge choices are possible. To briefly illustrate this, we
consider this field theory in the $\delta$-gauge, in which we work
directly with the functional integral in the form Eq.
(\ref{pathint2}) (and with the overall infinite term removed). Put
$\mathbf{f}(\mathbf{V}) = \mathbf{V_0}$ and $\mathbf{U} =
\mathbf{b}$ in Eq.(\ref{pathint2}), choose as the weight function
$G[\mathbf{U}] = 1$ rather than Eq.(\ref{popchoice}), and use the
representation
\begin{equation}
\delta^3({\bf V}_0 - {\bf b}) = \int \exp\left[i {\bf K} \cdot ({\bf
V}_0 -{\bf b})\right] \frac{d^3K}{(2 \pi)^3}.
\end{equation}
In this case, by using the final equality in Eq.(\ref{FP2}), for
expressing the determinant of the identity matrix as a Grassmann
integral, it gives
\begin{equation}\label{deltagauge}
S_{GF} = S[{\bf V}, {\bm \sigma}] - i {\bf K} \cdot ({\bf V}_0 -
{\bf b}) + i{\bm \eta}^* \cdot{\bm \eta}.
\end{equation}
The Galilean transformations are as above Eqs.(\ref{galZ11} -
\ref{galZ41}), for which $S[{\bf V}, {\bm \sigma}]$ and the measure
are invariant. For the BRS transformation, we identify ${\bf c}
\rightarrow \zeta ({\bm \eta}^* + {\bm \eta})$ as before with
\begin{eqnarray}\label{brsd1}
\delta_{\rm BRS} {\bf V}_{0j} & = & -{\zeta} ({\bf \eta}_j^* + {\bf \eta}_j),  \\
\label{brsd2} \delta_{\rm BRS} {\bf \eta}_j & = &   {\bf K}_j \zeta,
\\ \label{brsd3}
\delta_{\rm BRS} {\bf \eta}_j^* & = & - {\bf K}_j  {\zeta}, \\
\delta_{\rm BRS} \mathbf{b} &=& \delta_{\rm BRS} \mathbf{K} = 0.
\end{eqnarray}
We point out that ${\bm \eta},\bm\eta^*$ need not be a complex
conjugate pair; see footnote in Sec \ref{sec:BRSEGI}.  With this,
one can check
\begin{equation}
\delta_{BRS} \left( i{\bm \eta^*} \cdot {\bm \eta} - i {\bf K} \cdot
({\bf V}_0 - {\bf b}) \right) = 0 ,
\end{equation}
proving that the gauge-fixed action in this gauge
Eq.(\ref{deltagauge}) is BRS invariant: $\delta_{BRS}S_{GF} = 0$.

\subsection{\label{sec:boxward} Ward identity in a box and spurious relations}

In order to remove any ambiguity from our treatment of the continuum
NSE path integral in \cite{BH2} and the spurious relations we found
there, we here perform a similar analysis but now for a discrete and
hence regularized version of NSE field theory. In this Section we
use the dynamic functional $Z$ for the NSE derived in a box (see
Appendix \ref{sec:box}) to obtain the associated Ward identities. An
important technical distinction stems from the fact that in a box,
there is a smallest limiting finite velocity boost. Only in the
continuum can infinitesimal boosts be treated.

To begin, and including the sources from the outset, we start with
\begin{eqnarray}\label{discz}
Z[\mathbf{J},\mathbf{\Sigma}] &=& \int \prod_{\alpha,\mathbf{n},j}d
v_{\alpha}(\mathbf{n},j) d\sigma_{\alpha}(\mathbf{n},j)\times
\nonumber  \\
&&\exp \Big( -S[\mathbf{v},{\bm \sigma}] + \sum_{\mathbf{m},l}
\big(J_{\alpha}(-\mathbf{m},-l)v_{\alpha}(\mathbf{m},l) +
\Sigma_{\alpha}(-\mathbf{m},-l)\sigma_{\alpha}(\mathbf{m},l)\big)\Big).
\end{eqnarray}
Here, the discrete action $S$ is given by Eq.(\ref{discreteS}). We
repeat all the steps that for the continuum case, lead one to obtain
the Ward as for example in \cite{BH}. Here, we carry this out for
the discrete theory. So we next transform $Z$ using the discrete
version of the Galilean transformation Eq.(\ref{kgaldis}). Doing so
we obtain:
\begin{eqnarray}\label{zzz}
Z[\mathbf{J},\mathbf{\Sigma}] &=& \int \prod_{\alpha,\mathbf{n},j}d
v_{\alpha}(\mathbf{n},j) d\sigma_{\alpha}(\mathbf{n},j)\,\exp \Big(
-S[\mathbf{v},{\bm \sigma}] +
\sum_{\mathbf{m},l}\big(J_{\alpha}(-\mathbf{m},-l)v_{\alpha}(\mathbf{m},l
+ \bar \mathbf{c \cdot} \mathbf{m}) \nonumber \\
&-&
\bar\mathbf{c}\cdot\mathbf{J}(-\mathbf{m},-l)\delta^3(\mathbf{m})\delta(l)
+ \Sigma_{\alpha}(-\mathbf{m},-l)\sigma_{\alpha}(\mathbf{m},j + \bar
\mathbf{c \cdot} \mathbf{m})\big) \Big).
\end{eqnarray}
Now unlike the continuum case, we cannot consider an infinitesimal
boost velocity. Instead, the best we can do is consider the
\textit{smallest} nonzero boost velocity, which, according to
comments immediately below Eq.(\ref{kgaldis}), is $\delta \bar{c} =
(1,1,1)$. We could as well have chosen $(1,0,0)$, or $(0,1,0)$, or
$(0,0,1)$, or $(1,1,0)$, etc. The point is, the smallest allowable
boost velocity is \textit{not} infinitesimal.

We next emulate what is done for the continuum case, namely Taylor
expand the velocity and conjugate fields about zero boost, see for
example, Eq.(4) in \cite{BH}. The best we can do is to employ the
following approximation:
\begin{eqnarray}
v_{\alpha}(\mathbf{m},l + \bar{\mathbf{c}}\cdot\mathbf{m}) &=&
v_{\alpha}(\mathbf{m},l) + \bar{\mathbf{c}}\cdot\mathbf{m}\Delta
v_{\alpha}(\mathbf{m},l)\nonumber \\
&=&  v_{\alpha}(\mathbf{m},l) + (m_1+m_2+m_3)\Delta
v_{\alpha}(\mathbf{m},l),
\end{eqnarray}
where the finite difference operator $\Delta f(\mathbf{m},l) =
f(\mathbf{m},l+1)-f(\mathbf{m},l)$.

To finish the derivation, we insert these finite difference
approximations back into Eq.(\ref{zzz}). This implies that the
functional transforms as $Z \rightarrow Z + \delta Z$. We require
that the extra terms proportional to the boost velocity
$\bar{\mathbf{c}}$ vanish identically. This leads to:
\begin{eqnarray}\label{discreteWT}
\sum_{\mathbf{m},l}&\Big(& J_{\alpha}(-\mathbf{m},-l)m_i \Delta
\frac{\partial}{\partial J_{\alpha}(-\mathbf{m},-l)} \nonumber \\
&+& \Sigma_{\alpha}(-\mathbf{m},-l)m_i \Delta
\frac{\partial}{\partial \Sigma_{\alpha}(-\mathbf{m},-l)} -
J_i(-\mathbf{m},-l)\delta^3_{\mathbf{m},\mathbf{0}}\delta_{l,0}\Big)Z[\mathbf{J},\mathbf{\Sigma}]
= 0.
\end{eqnarray}
This is the Ward identity for $Z$ and should be compared to the
Fourier transform of the continuum version Eq.(6) in \cite{BH}.

Differentiate this WTI for $Z$, Eq.(\ref{discreteWT}), with respect
to $\frac{\partial }{\partial J(-\mathbf{m},-l)}|_{J=\Sigma=0} $, to
obtain
\begin{equation}\label{spur}
m_i\Delta <v_k(\mathbf{m},l)> -  \delta_{ik}= 0.
\end{equation}
This is well behaved and never singular, but it suffers from the
same defect as its continuum version \cite{BH2} when we set $i = k$
and sum over this index: fluid incompressibility implies
$m_iv_i(\mathbf{m},l) = 0$, which since it is zero, leads to a
contradiction in Eq.(\ref{spur}).

For higher point correlations, more spurious relations are found
from Eq.(\ref{discreteWT}), similar to those found in the continuum
case in \cite{BH2}. The source of these spurious relations is the
same as in \cite{BH2}, and we can verify that the functional
integral in Eq.(\ref{zzz}) has an infinite prefactor due to the
integration over the velocity zero mode. This can be fixed by the
same gauge fixing procedure as we carried out in \cite{BH2}, taking
appropriate modifications for the discrete case. Most importantly,
the treatment in this Section demonstrates that these spurious
relations are not an artifact of the continuum path integral
expressions.

\section{\label{sec:egi} Extended Galilean Invariance (EGI): Gauge fixing and BRS Symmetry}

Galilean invariance admits an interesting physical extension.  As it
turns out, the Navier-Stokes (NSE) equation is invariant under
rectilinear frame accelerations, a symmetry known as extended
Galilean invariance (EGI) \cite{Pope}.  This invariance was noted
and used in the context of the KPZ equation \cite{LL}, as well as in
the stochastic Burgers equation \cite{Ivash}, it has been used in
probability density functions in turbulence modeling \cite{Tong},
and finds practical applications in fluid animation simulations
\cite{Shah}. We review this invariance of the NSE below. We then
proceed to gauge-fix this symmetry, write down the corresponding BRS
transformation that leaves the gauge-fixed action invariant, and
then deduce the associated Slavnov-Taylor identity for the
functional and the new generalized Ward identity that follows from
it.

\subsection{\label{sec:egiNSE} Extended Galilean invariance of the Navier-Stokes equation}

Consider the Navier-Stokes equation (NSE) in an inertial frame $E$:
\begin{equation}
\label{NSE} \frac{\partial V_{i}}{\partial t} + \frac{\partial
(V_{i}V_{j})}{\partial x_{j}} = -\frac{\partial \Pi}{\partial x_{i}}
+ \nu \nabla^{2}V_{i},
\end{equation}
where $\nu$ is the kinematic viscosity of the fluid, $V_{i}({\bf x},
t)$ and $\Pi({\bf x}, t)$ are the instantaneous values of the
velocity and pressure, and the continuity equation takes the form
\begin{equation}
\label{conti} \frac{\partial V_i}{\partial x_i} =0,
\end{equation}
for an incompressible fluid. In this case, the density is constant
for all $\mathbf{x}$ and $t$ so for convenience we may work in a
system of units where the fluid density is taken to be unity. Also,
as is well known, taking the divergence of each term in
Eq.(\ref{NSE}), and invoking Eq.(\ref{conti}), leads to a
Poisson-type equation for the pressure, viz.,
\begin{equation} \label{press}
\nabla^2 \Pi = -\frac{\partial^2 (V_i V_j)}{\partial x_i \partial
x_j},
\end{equation}
and this result is useful for establishing the Galilean
transformation of the fluid pressure.

Now consider a second \textit{noninertial} frame $\bar E$ moving
with respect to $E$ with a variable but rectilinear velocity
$\mathbf{c}(t)$. The transformation between these two coordinate
systems is as follows,
\begin{eqnarray}\label{exgaltrans1}
\mathbf{x}' &=& \mathbf{x} - \bm{\lambda}(t), \\\label{exgaltrans2}
t' &=& t , \\\label{exgaltrans3} \mathbf{V}'(\mathbf{x}',t') &=&
\mathbf{V}(\mathbf{x},t) - \bm{\dot\lambda}(t),
\\\label{exgaltrans4} \Pi'(\mathbf{x}',t') &=& \Pi(\mathbf{x},t) +
\mathbf{x}'\cdot \bm{\ddot\lambda}(t),
\end{eqnarray}
where $\bm{\lambda}(t) = \int_0^t \, \mathbf {c}(s) ds$, and the
dots stand for the derivative taken with respect to time. Applying
these transformation rules to (\ref{NSE}) proves that the
transformed NSE takes the form

\begin{equation}\label{accelNSE}
\label{NSE'} \frac{\partial V'_{i}}{\partial t'} + \frac{\partial
(V'_{i}V'_{j})}{\partial x'_{j}} = -\frac{\partial \Pi'}{\partial
x'_{i}} + \nu \nabla'^{2}V'_{i},
\end{equation}
which is the NSE in the noninertial \textit{accelerating} frame.
Note that $\bm{\ddot\lambda}(t) = \dot \mathbf {c}(t) =
\mathbf{A}(t)$, so that the frame acceleration is absorbed into a
modified pressure Eq.(\ref{exgaltrans4}). This demonstrates that the
NSE Eq.(\ref{NSE}) is invariant under rectilinear frame
accelerations. It is clear that when the above transformation rules
Eqs.(\ref{exgaltrans1}-\ref{exgaltrans4}) include one for the
conjugate field:
\begin{equation}
\label{exgaltrans5} \bm{\sigma}'(\mathbf{x}',t') =
\bm{\sigma}(\mathbf{x},t),
\end{equation}
then the action Eq.(\ref{class}) and dynamic functional Eq.(\ref{Z})
are invariant as well under this extended Galilean transformation.

\subsection{\label{sec:FPdetEGI} Gauge Fixing for Extended Galilean invariance}

We next gauge-fix the EGI of the dynamic functional in Eq.(\ref{Z}).
Once again, we refer to the general gauge-fixing procedure as
outlined in Appendix \ref{appendix}. We make the specific gauge
choice introduced below, calculate the corresponding Fadeev-Popov
determinant Eq.(\ref{FPdet}), and once again select the arbitrary
function $\mathbf{U}$ that appears in Eq.(\ref{pathint2}) to
complete the procedure.

In the case of EGI, the purpose of gauge fixing is to single out one
rectilinearly \textit{accelerating} reference frame. This can be
accomplished by now constraining the time-dependent \textit{zero
mode} of the full instantaneous velocity field. This mode
corresponds to the acceleration of the system as a whole, that is,
its bulk acceleration.  We therefore make the following choice for
the gauge function, where $vol = L^3$ is the volume of the spatially
bounded box:
\begin{equation}\label{EGIgauge}
\mathbf{f}(\mathbf{V}) = \frac{1}{vol}\int d\mathbf{x} \,
\mathbf{V}(\mathbf{x},t) \equiv \mathbf{V_0}(t).
\end{equation}
This projects out the instantaneous and arbitrary time dependent
bulk velocity of the entire bounded system as required.  From
Eqs.(\ref{exgaltrans1}-\ref{exgaltrans3}) we have
\begin{eqnarray}\label{vlam1}
V_j^{\bm{\lambda}}(\mathbf{x},t) &=& V'_j(\mathbf{x'},t') +
\dot\lambda_j(t), \\ \label{vlam2} &=&
V'_j(\mathbf{x}-\bm{\lambda}(t),t) + \dot\lambda_j(t),
\\\label{vlam3} &=& V'_j(\mathbf{x},t)
-\lambda_k(t)\frac{\partial}{\partial x_k}V'_j(\mathbf{x},t) +
\dot\lambda_j(t) + O(\lambda^2), \\ \label{vlam4}{\rm So}\,\,\,
\frac{\delta V_j^{\bm\lambda}(\mathbf{x},t)}{\delta
\lambda_m(t')}|_{\bm{\lambda = 0}} &=& \Big(\frac{\partial
V'_j(\mathbf{x},t)}{\partial x_m} -
\delta_{jm}\frac{d}{dt}\Big)\delta(t-t').
\end{eqnarray}
The FP determinant Eq.(\ref{FPdet}) in this case works out to be,
again using periodic boundary conditions for the bounded spatial
box,
\begin{eqnarray}\label{FP4}
\Delta_f[\mathbf{V}] = \det \Big( \frac{\delta
f_i(t;\mathbf{V}^{\lambda})}{\delta \lambda_m(t')}\Big)|_{\bm\lambda
= \mathbf{0}} &=& \det(-\delta_{im}\frac{d}{dt}\delta(t-t')) \nonumber \\
&=& \int {\cal D}\bm{\eta} {\cal D}\bm{\eta}^* \exp\Big(-i\int dt\,
\bm{\eta}(t)^* \cdot\frac{d}{dt} \bm{\eta}(t) \Big).
\end{eqnarray}
The functional integral in Eq.(\ref{FP4}) is over two time-dependent
Grassman vector fields $\bm{\eta}(t)$ and $\bm{\eta}^*(t)$.   For
the convergent integral over the gauge group volume, we now choose
(see Eq.(\ref{popchoice}) in Appendix {\ref{appendix})
\begin{equation}\label{popEGI}
G[\mathbf{b}] = \exp\Big( -\frac{1}{2\xi}\int dt \,
\mathbf{b}^2(t)\Big),
\end{equation}
which yields the gauge-fixed action for EGI:
\begin{equation}\label{GFactionEGI}
S_{GF} = S[\mathbf{V},\bm{\sigma}] + \int dt\,\left\{ \frac{1}{2\xi}
\mathbf{V_0}^2(t) + i\bm{\eta^*}(t)\frac{d}{dt}\bm{\eta}(t)\right\}.
\end{equation}
At this point, Extended Galilean invariance is manifestly broken in
$S_{GF}$.

\subsection{\label{sec:BRSEGI} Becchi-Rouet-Stora transformation for gauge fixed EGI}

We begin by writing down the BRS transformation. We observe that the
transformations in Eqs.(\ref{EGIBRS1}-\ref{EGIBRS5}) below are just
the \textit{infinitesimal} version of the extended Galilean
transformations Eqs.(\ref{exgaltrans1}-\ref{exgaltrans4}), but with
the variable rectilinear boost velocity now replaced by
$\bm{\dot\lambda}(t) = \mathbf{c}(t) \rightarrow \zeta
\bm{\dot{\eta}}(t)$, where $\zeta$ is a real constant Grassmann
parameter: $\zeta^2 = 0$. Therefore the nongauge-fixed action $S$
and the path integral measure are automatically invariant under this
subset of BRS transformations:
\begin{eqnarray}\label{EGIBRS1}
\delta_{\rm BRS}\mathbf{x} &=& -\zeta \bm{\eta}(t),
\\\label{EGIBRS2} \delta_{\rm BRS} t &=& 0 ,\\\label{EGIBRS3}
\delta_{\rm BRS} \mathbf{V}(\mathbf{x},t) &=& \zeta \eta_k(t)
\frac{\partial \mathbf{V}(\mathbf{x},t)}{\partial x_k} - \zeta
\bm{\dot \eta}(t),
\\\label{EGIBRS4} \delta_{\rm BRS} \bm{\sigma}(\mathbf{x},t) &=&
\zeta \eta_k(t) \frac{\partial \bm{\sigma}(\mathbf{x},t)}{\partial
x_k},
\\\label{EGIBRS5}
\delta_{\rm BRS}\Pi(\mathbf{x},t) &=& \zeta \eta_k(t) \frac{\partial
\Pi(\mathbf{x},t)}{\partial x_k} +
\mathbf{x}\cdot\zeta\bm{\ddot{\eta}}(t).
\end{eqnarray}
To complete the BRS transformation, we need to know how the velocity
zero mode $\mathbf{V_0}(t)$ and Grassmann sector transform. The
remaining BRS transformations are given as follows:
\begin{eqnarray}\label{EGIBRS6}
\delta_{\rm BRS}\mathbf{V_0}(t) &=& -\zeta\bm{\dot\eta}(t), \\
\label{EGIBRS7} \delta_{\rm BRS}\bm{\eta}(t) &=& 0
,\\\label{EGIBRS8} \delta_{\rm BRS}\bm{\eta^*}(t) &=&
\frac{-i}{\xi}\mathbf{V_0}(t)\zeta.
\end{eqnarray}
Note that Eq.(\ref{EGIBRS6}) is an immediate consequence of
Eq.(\ref{EGIBRS3}). Secondly, note the rather different ways that
$\bm{\eta}$ and $\bm{\eta^*}$ transform. This ``asymmetry" is in
fact familiar from non-abelian quantum field theory \footnote{See
for example the chapter on ``Yang-Mills Theory: Slavnov-Taylor
Identities", page 370 in \cite{Ramond} for full details of the proof
of the BRS invariance of gauge-fixing and ghost terms in a general
Yang-Mills gauge theory. Note also the use there of two independent
real Grassmann fields. The limit of vanishing Lie algebra structure
constants (the Abelian limit) in Yang-Mills quantum field theory
implies that one of the two ghost fields transforms to zero under
BRS.  See also remarks \cite{Pokorski} to the effect that
$\bm{\eta}$ and $\bm{\eta^*}$ can be independent, it is not
necessary to think of them as hermitian conjugates, indeed they can
be any two independent anticommuting fields. }

Summarizing up to this point, we have demonstrated that the gauge
fixed action Eq.(\ref{GFactionEGI}) is invariant under the above BRS
transformation:
\begin{equation}
\delta_{\rm BRS}S_{GF} = 0.
\end{equation}
%

\subsection{\label{sec:WTIEGI} The Generalized Slavnov-Taylor and Ward identities}

To conclude, we subject $Z_{GF}$, which we now define below, to the
BRS transformation and derive the generalized Ward identity that
holds for $\Gamma$,
\begin{eqnarray}\label{gfZEGI}
Z_{GF}[\mathbf{J},\bm{\Sigma},\bm{\theta},\bm{\theta}^*] &=& \int
[D{\bf V}][D \bm{\sigma}][D\bm{\eta}][D\bm{\eta}^*]
\exp\{-S_{GF}[{\bf V}, \bm{\sigma},
\bm{\eta},\bm{\eta}^*] + \int dt \, (\bm{\theta}^*\cdot\bm{\eta} + \bm{\theta}\cdot\bm{\eta}^*) \nonumber \\
&+& \int d\mathbf{x} dt \,
\{\mathbf{J}(\mathbf{x},t)\cdot\mathbf{V}(\mathbf{x},t) +
\mathbf{\Sigma}(\mathbf{x},t)\cdot\bm{\sigma}(\mathbf{x},t) \} \}.
\end{eqnarray}
Here of course, the gauge fixed action $S_{GF}$ is given by
Eq.(\ref{GFactionEGI}), and the Grassmann vector sources
$\bm{\theta}(t),\bm{\theta}^*(t)$ are now time dependent. We next
subject this functional to the complete BRS transformation listed in
Eqs.(\ref{EGIBRS1}-\ref{EGIBRS8}). Just as in the case treated in
Sec \ref{sec:ST}, the measure and gauge-fixed action are invariant
and only the source terms will be affected. The remainder of the
steps involved are similar to those employed above. We expand the
exponential to first order in $\zeta$, and then make use of the
Legendre transform similar to Eq.(\ref{gamma0}) to define the
effective action $\Gamma$. For the present case, the Grassmann
source terms are integrated over time in the Legendre transform.
After a few simple steps we arrive at the following Slavnov-Taylor
identity:
\begin{equation}\label{EGIidentity}
\left[\frac{i}{\xi} \int dt \,\frac{\delta\Gamma}{\delta
{\bm{\eta}^*_{cl}(t)}} \cdot \mathbf{V_0}^{cl}(t)  + \int
d\mathbf{x} \int dt\, \eta^{cl}_j(t)\Big(\frac{\delta \Gamma}{\delta
V^{cl}_l}   \nabla_j V^{cl}_l + \frac{\delta \Gamma}{\delta
\sigma^{cl}_l} \nabla_j \sigma^{cl}_l + \frac{\partial}{\partial
t}\frac{\delta \Gamma}{\delta V^{cl}_j} \Big) \right] = 0.
\end{equation}
The pertinent Ward identity we seek is obtained by inserting the
functional Taylor series for $\Gamma$ (see e.g., Eq.(9,10) of
Ref.\cite{BH}) into Eq.(\ref{EGIidentity}), differentiating the
latter with respect to $V_k(\mathbf{y},t')$ and
$\sigma(\mathbf{w},t'')$, and then setting all the fields equal to
zero. We immediately see that term depending on the gauge parameter
$\xi$ will not survive this sequence of steps and furthermore, since
$\bm{\eta}^{cl}(t)$ is an \textit{arbitrary} function of time, the
identity Eq.(\ref{EGIidentity}) effectively reduces to
\begin{equation}\label{WTI}
\int d^d\mathbf{x}\, \Big( \nabla_j V_l(\mathbf{x},t) \frac{\delta
\Gamma}{\delta V_l(\mathbf{x},t)} + \nabla_j \sigma_l(\mathbf{x},t)
\frac{\delta \Gamma}{\delta \sigma_l(\mathbf{x},t)} + \frac{\partial
}{\partial t}\frac{\delta \Gamma}{\delta V_j(\mathbf{x},t)}\Big) =
0.
\end{equation}
Carrying out the specific differentiations mentioned above yields a
relation between the exact inverse response $\Gamma^{(1,1)}$ and the
vertex functions $\Gamma^{(2,1)}$:
\begin{eqnarray}\label{WTIb}
\frac{\partial}{\partial t}\int
d^d\mathbf{x}\,\Gamma^{(2,1)}_{nkj}(\mathbf{x},t;\mathbf{y},t';\mathbf{w},t'')
&=& \delta(t-t'')\frac{\partial}{\partial
w_n}\Gamma^{(1,1)}_{kj}(\mathbf{y},t';\mathbf{w},t)\nonumber \\
&+& \delta(t-t')\frac{\partial}{\partial
y_n}\Gamma^{(1,1)}_{kj}(\mathbf{y},t;\mathbf{w},t'').
\end{eqnarray}
This is the new Ward identity that follows from EGI in configuration
space and time, derived here for the first time.  Whereas the vertex
function in the conventional Ward identity (see, e.g., Eq.(11) of
Ref.\cite{BH}) is integrated over both coordinates and time, here by
contrast, the vertex is integrated only over coordinates and is
differentiated with respect to time. Using translational invariance
(in space and in time), Fourier transforming (\ref{WTIb}) and then
integrating over $\int dt\int dt'\int dt'' \, e^{-i\omega_1
t}e^{-i\omega_2 t'}e^{-i\omega_3 t''}$ to get rid of the
delta-functions, we obtain the Ward identity in wavevector and
frequency space (note: here, $\nu$ stands for a frequency, not the
viscosity):
\begin{equation}\label{WTIc}
k_n\Gamma^{(1,1)}_{kj}( \mathbf{k},\omega + \nu; -\mathbf{k},-\omega
- \nu ) - k_n\Gamma^{(1,1)}_{kj}(\mathbf{k},\nu; -\mathbf{k},-\nu) =
-\omega \Gamma^{(2,1)}_{nkj}(\mathbf{0},\omega;\mathbf{k},\nu;
-\mathbf{k},\omega -\nu).
\end{equation}
If we take the \textit{zero frequency limit} $\omega \rightarrow 0$,
we then obtain the identity
\begin{equation}\label{limitWTI}
k_n \frac{\partial }{\partial \nu}\Gamma_{kj}^{(1,1)}(
\mathbf{k},\nu; -\mathbf{k},-\nu) =
-\Gamma^{(2,1)}_{nkj}(\mathbf{0},0;\mathbf{k},\nu;
-\mathbf{k},-\nu),
\end{equation}
which is precisely the Ward identity Eq.(\ref{WTIk2}) that follows
from the standard Galilean invariance of the NSE, that is, when the
frame $\bar E$ moves with a \textit{constant} velocity
$\bm{\ddot\lambda}(t) = \mathbf{0}$ with respect to the lab frame
$E$ \cite{BH}.

For EGI, there is second way to implement the BRS transformation,
intimately related to how we choose to define the gauge-function.
Thus, if we now choose $\bm{\dot \lambda}(t)$ \textit{instead} of
$\bm \lambda (t)$ as the gauge function, then in place of
Eq.(\ref{vlam4}), we would have, after using the identity
$\lambda_k(t) = \int_0^t du\frac{d\lambda_k(u)}{du} = \int_0^t du \,
\dot \lambda_k(u)$, the relation
\begin{equation}
\frac{\delta V_j^{\bm\lambda}(\mathbf{x},t)}{\delta \dot
\lambda_m(t')}|_{\bm{\dot \lambda = 0}} = \frac{\partial
V'_j(\mathbf{x},t)}{\partial x_m} - \delta_{jm}\,\delta(t-t').
\end{equation}
The FP determinant, invoking the same boundary conditions as before,
is now given by
\begin{eqnarray}
\Delta_f[\mathbf{V}] = \det \Big( \frac{\delta
f_i(t;\mathbf{V}^{\lambda})}{\delta \dot \lambda_m(t')}\Big)|_{\bm
\dot \lambda = \mathbf{0}} &=& \det(-\delta _{im}\delta(t-t'))
\nonumber \\
&=& \int {\cal D}\bm{\eta} {\cal D}\bm{\eta}^*
\exp\Big(-i\int dt\, \bm{\eta}(t)^* \cdot \bm{\eta}(t) \Big),
\end{eqnarray}
and then finally Eq.(\ref{popEGI}) yields the gauge fixed action
\begin{equation}\label{GFactionEGI2}
S_{GF} = S[\mathbf{V},\bm{\sigma}] + \int dt\,\left\{ \frac{1}{2\xi}
\mathbf{V_0}^2(t) + i\bm{\eta^*}(t)\cdot\bm{\eta}(t)\right\},
\end{equation}
to be contrasted with Eq.(\ref{GFactionEGI}). The net result of this
alternative choice of gauge function is to \textit{remove the time
derivative} from the Grassmann fields in the gauge fixed action
Eq.(\ref{GFactionEGI2}).

In this latter case, the BRS transformation that leaves the
gauge-fixed action Eq.(\ref{GFactionEGI2}) invariant includes in
part the infinitesimal extended Galilean transformation obtained by
replacing the variable boost velocity by the time dependent
Grassmann vector $\mathbf{c}(t) \rightarrow \zeta \bm{\eta}(t)$,
where $\zeta$ is the real Grassmann constant we introduced before.
Namely,
\begin{eqnarray}\label{EGIBRSa}
\delta_{\rm BRS}\mathbf{x} &=& -\zeta \int_0^t du \, \bm{\eta}(u),
\\\label{EGIBRSb} \delta_{\rm BRS} t &=& 0 ,\\\label{EGIBRSc}
\delta_{\rm BRS} \mathbf{V}(\mathbf{x},t) &=& \zeta \int_0^t du \,
\bm{\eta}_k (u) \frac{\partial \mathbf{V}(\mathbf{x},t)}{\partial
x_k} - \zeta \bm{\eta}(t),
\\\label{EGIBRSd} \delta_{\rm BRS} \bm{\sigma}(\mathbf{x},t) &=&
\zeta \int_0^t du \, \bm{\eta}_k (u) \frac{\partial
\bm{\sigma}(\mathbf{x},t)}{\partial x_k},
\\\label{EGIBRSe}
\delta_{\rm BRS}\Pi(\mathbf{x},t) &=& \zeta \int_0^t du \,
\bm{\eta}_k (u) \frac{\partial \Pi(\mathbf{x},t)}{\partial x_k} +
\mathbf{x}\cdot\zeta\bm{\dot{\eta}}(t).
\end{eqnarray}
This version of the BRS transformation reduces to that found in
\cite{BH2} in the limit that EGI goes over to standard GI (zero
frame acceleration). To complete the BRS transformation, we must
specify how the velocity zero mode and Grassmann variables
transform. These are simply
\begin{eqnarray}\label{EGIBRSg}
\delta_{\rm BRS}\mathbf{V_0}(t) &=& -\zeta\bm{\eta}(t), \\
\label{EGIBRS7} \delta_{\rm BRS}\bm{\eta}(t) &=& 0
,\\\label{EGIBRSh} \delta_{\rm BRS}\bm{\eta^*}(t) &=&
\frac{-i}{\xi}\mathbf{V_0}(t)\zeta.
\end{eqnarray}
Note: we could also choose the more ``symmetric" assignment
$\mathbf{c}(t) \rightarrow \zeta (\bm{\eta}(t) + \bm{\dot
\eta}(t))$, and then make the associated simple adjustments to the
above set of transformation rules, but this has no effect upon the
generalized Ward identity that follows from BRS. To summarize so
far, the gauge fixed action Eq.(\ref{GFactionEGI2}) is invariant
under this BRS transformation Eqs(\ref{EGIBRSa}-\ref{EGIBRSh}).

To finish, we next subject $Z_{GF}$ in Eq.(\ref{gfZEGI}) to the
above BRS transformation, where now $S_{GF}$ is given by
Eq(\ref{GFactionEGI2}). Following the now rather familiar steps,
this leads to the equation $\delta_{BRS} Z_{GF} = 0$, and expressing
this in terms of the effective action $\Gamma$, yields the
Slavnov-Taylor identity
\begin{equation}\label{secondEGIidentity}
\left[\frac{i}{\xi} \int dt \,\frac{\delta\Gamma}{\delta
{\bm{\eta}^*_{cl}(t)}} \cdot \mathbf{V_0}^{cl}(t)  + \int
d\mathbf{x} \int dt\, \rho^{cl}_j(t)\Big(\frac{\delta \Gamma}{\delta
V^{cl}_l}   \nabla_j V^{cl}_l + \frac{\delta \Gamma}{\delta
\sigma^{cl}_l} \nabla_j \sigma^{cl}_l + \frac{\partial}{\partial
t}\frac{\delta \Gamma}{\delta V^{cl}_j} \Big) \right] = 0,
\end{equation}
where we have defined $\bm{\rho}^{cl}(t) = \int_0^t du\,
\bm{\eta}^{cl}(u)$, so that $\bm{\dot \rho}^{cl}(t) =
\bm{\eta}^{cl}(t)$. The Slavnov-Taylor identities that follow from
EGI  Eqs.(\ref{EGIidentity},\ref{secondEGIidentity}) and the Ward
identities that they imply Eqs.(\ref{WTI}-\ref{WTIc}) have not, to
our knowledge, been derived nor considered before. From this point
on, the entire discussion immediately following
Eq.(\ref{EGIidentity}) remains intact and applies without
modification to Eq.(\ref{secondEGIidentity}), as does the discussion
concerning the implications of this identity
Eq.(\ref{secondEGIidentity}) for vertex renormalization.

\section{\label{sec:QED} Ward identities in QED and in the stochastic field theory of randomly stirred fluids}

The well-known Ward identity in QED relating the exact
photon-electron vertex and the electron propagator reads,
\begin{equation}\label{QEDWT}
{S_F'}^{-1}(p+q) - {S_F'}^{-1}(p) = q^{\mu}\Gamma_{\mu}(p+q,p),
\end{equation}
where $p+q,q$ denote the four-momentum of the entering and emerging
electron lines, respectively, $\Gamma_\mu$ denotes the exact
photon-fermion-fermion vertex and ${S_F'}(k)$ is the exact Feynman
fermion propagator \cite{BD}. This identity allows us to compute
$S_F'$ directly from knowledge of the vertex and is represented
graphically in Eq.(\ref{QEDWTdiag}) where the blobs denote the exact
expressions. The four-momenta carried by the individual fermion
lines (single arrowed-lines) is indicated by the quantities within
parentheses. Note that the four-momenta carried by the photon line
(wiggly) on the right-hand side of Eq.(\ref{QEDWTdiag}) is therefore
$= q$. This identity is obviously satisfied by the bare vertex
$\Gamma_{\mu} = \gamma_{\mu}$ and bare propagator, $S_F(p) = (
p_{\mu}\gamma^{\mu} - m)^{-1}$.

\begin{fmffile}{fmfdgram24}
\begin{equation}\label{QEDWTdiag}
(p+q)\,\parbox{40mm}{\begin{fmfchar*}(35,35) \fmfleft{i1}
\fmfright{o1} \fmf{plain}{i1,v1} \fmf{fermion}{v1,o1}
\fmfblob{.15w}{v1}
\end{fmfchar*}} \,\, -
\,(p)\parbox{40mm}{\begin{fmfchar*}(35,35) \fmfleft{i1}
\fmfright{o1} \fmf{plain}{i1,v1} \fmf{fermion}{v1,o1}
\fmfblob{.15w}{v1}
\end{fmfchar*}} \,\,
 = \,(p+q) \,\parbox{40mm}{\begin{fmfchar*}(35,35) \fmfleft{i1} \fmftop{ti}
\fmfright{o1} \fmf{fermion}{i1,v1} \fmf{fermion}{v1,o1} \fmffreeze
\fmf{photon}{v1,ti} \fmfblob{.15w}{v1}
\end{fmfchar*}}\,(p)
\end{equation}
Taking the $q\rightarrow 0$ limit of Eq.(\ref{QEDWT}) yields
\begin{equation}\label{QEDWard}
\frac{\partial {S_F'}^{-1}(p)}{\partial p^{\mu}} =
\Gamma_{\mu}(p,p),
\end{equation}
which holds for the special case when the \textit{momentum transfer}
to the photon approaches zero. The crux of this identity is the
observation that the derivative of a fermion line with respect to
the four-momentum is equivalent to the insertion of a photon line in
the limit of zero momentum transfer. This derivative rule therefore
generates the ``soft photon" vertex in QED. We illustrate this
differential identity Eq.(\ref{QEDWard}) diagramatically in
Eq.(\ref{QEDWdiag}): note that the four-momenta of the photon line
(wiggly) is zero: $q=0$.
\begin{equation}\label{QEDWdiag}
\frac{\partial}{\partial p_{\mu}}\{\,\,
\parbox{40mm}{\begin{fmfchar*}(40,35)
\fmfleft{i1} \fmfright{o1} \fmf{plain}{i1,v1} \fmf{fermion}{v1,o1}
\fmfblob{.15w}{v1} \fmffreeze
\end{fmfchar*}}\,\, (p) \}\,\,
 = (p)\,\parbox{40mm}{\begin{fmfchar*}(40,35) \fmfleft{i1} \fmftop{ti}
\fmfright{o1} \fmf{fermion}{i1,v1} \fmf{fermion}{v1,o1} \fmffreeze
\fmf{photon}{v1,ti} \fmfblob{.15w}{v1}
\end{fmfchar*}}\, (p)
\end{equation}
Having briefly reviewed the main features of the key Ward identity
in QED \cite{BD}, we come back to the analogous identities that hold
in the stochastic field theory of the randomly forced NSE. The exact
identity Eq.(\ref{WTIc}) is shown in diagram form in
Eq.(\ref{NSEWTdiag}) and holds for the exact inverse response
function and triple vertex (indicated with blobs). The zig-zag and
arrowed double lines denote the conjugate and velocity fields,
respectively. The momentum and frequency carried by the various
lines is indicated by the associated pair of variables $({\bm
k},\omega)$. Note that the momentum and frequency carried by the
vertical velocity line (arrowed double line) on the right-hand side
of Eq.(\ref{NSEWTdiag}) is therefore $= (\bm{0},\omega)$. This
identity is obviously satisfied by the bare inverse response
function $\Gamma_{ij}^{(1,1)}(\bm{k},\omega) = (-i\omega + \nu_0
k^2)P_{ij}(\bm{k})$ and bare vertex $\Gamma^{(2,1)}_{nij} = ik_n
P_{ij}(\bm{k})$ \cite{Mou}. This identity is structurally quite
similar to the QED Ward identity in (\ref{QEDWT}) but there is an
important difference: in Eq.(\ref{WTIc}) the 3-momentum transfer to
the inserted velocity line is always zero, whereas any finite
frequency $\omega$ can be transferred.
\begin{eqnarray}\label{NSEWTdiag}&k_n&\left\{(\bm{k},\omega+\nu)\,
\parbox{40mm}{\begin{fmfchar*}(40,35)
\fmfleft{i1} \fmfright{o1} \fmf{zigzag}{i1,v1} \fmf{heavy}{v1,o1}
\fmfblob{.15w}{v1} \end{fmfchar*}} \,-  (\bm{k},\nu)\,
\parbox{40mm}{\begin{fmfchar*}(40,35) \fmfleft{i1} \fmfright{o1}
\fmf{zigzag}{i1,v1} \fmf{heavy}{v1,o1} \fmfblob{.15w}{v1}
\end{fmfchar*}} \right\} \nonumber \\
&=& -\omega \{(\bm{k},\omega+\nu)\,
\parbox{40mm}{\begin{fmfchar*}(40,35) \fmfleft{i1} \fmftop{ti}
\fmfright{o1} \fmf{zigzag}{i1,v1} \fmf{heavy}{v1,o1} \fmffreeze
\fmf{heavy}{v1,ti} \fmfblob{.15w}{v1}
\end{fmfchar*} }\,(\bm{k},\nu)\}
\end{eqnarray}
Taking the zero-frequency limit ($\omega \rightarrow 0$) yields the
further differential identity in Eq.(\ref{limitWTI}), which as we
see, holds for the special case when the \textit{frequency transfer}
to the inserted zero-momentum velocity line goes to zero. The crux
of this identity is the simple observation that the derivative of
the exact response function with respect to frequency is equivalent
to the insertion of a velocity line in the limit of \textit{both}
vanishing momentum and frequency transfer: $= (\bm{0},0)$.  We
illustrate this differential identity in Eq.(\ref{NSEWdiag}). This
derivative operation therefore generates the ``\textit{soft
velocity}" vertex in NSE diagrammatic perturbation theory: the
momentum-frequency pair carried by the velocity (the vertical
arrowed double line on the right-hand side of Eq.(\ref{NSEWdiag}))
is $= (\bm{0},0)$.
\begin{equation}\label{NSEWdiag}
-k_n\frac{\partial}{\partial \nu}\{\,\,
\parbox{40mm}{\begin{fmfchar*}(40,35)
\fmfleft{i1} \fmfright{o1} \fmf{zigzag}{i1,v1} \fmf{heavy}{v1,o1}
\fmfblob{.15w}{v1}
\end{fmfchar*}} \,(\bm{k},\nu)\}\,\,
= (\bm{k},\nu)\,\, \parbox{40mm}{\begin{fmfchar*}(40,35)
\fmfleft{i1} \fmftop{ti} \fmfright{o1} \fmf{zigzag}{i1,v1}
\fmf{fermion}{v1,o1} \fmffreeze \fmf{fermion}{v1,ti}
\fmfblob{.15w}{v1}
\end{fmfchar*}}\,(\bm{k},\nu)
\end{equation}
\end{fmffile}

The rigid frame velocity, relative to the lab frame, is what enters
into the WTI's for the NSE. This velocity can either be a constant,
in which case we have the strict Galilean transformation, or it can
be time dependent but rectilinear, in which case we have the
extended Galilean transformation. This velocity enters the triple
vertex as a \textit{background} field. This is further supported by
looking directly at the NSE in momentum space and time \cite{McBook}
and examining the nonlinear contribution. This contribution sums
over all velocity modes and time dependences; there is a single
contribution coming from the time dependent zero-mode which is
simply the contribution coming from rigidly translating the
reference frame. So the WTI is the statement that it is only this
rigid motion that does not renormalize. This bulk system motion can
correspond to a constant velocity, or a time dependent but
rectilinear velocity, i.e., an arbitrary rectilinear acceleration.

Regarding our comparison to gauge field theory, this has a two-fold
motivation. On the one hand, we note that the fluid velocity
transformation $\bf V \rightarrow V' = V + c$ and the transformation
of the QED vector potential $\bf A \rightarrow A' = A + \nabla \phi$
are affine transformations leaving the convective derivative in the
Navier-Stokes equation invariant and the $U(1)$ covariant derivative
invariant, respectively. The second motivation is inspired in large
part by the structure of the Ward identities that follow from
breaking Galilean (or extended Galilean) invariance, since these
identities Eqs.(\ref{WTIc},\ref{limitWTI}) do closely resemble those
that follow from breaking gauge invariance in QED,
Eqs.(\ref{QEDWT},\ref{QEDWard}), respectively, as we argued above.
QED of course has a local (gauge) invariance, whereas Galilean
invariance is a global symmetry.

One can also make an analogy between Galilean invariance and field
theories with global symmetries. For example consider symmetry
breaking in a globally invariant field theory such as N-component
scalar field theory with a global $O(N)$ invariance, where many
results have been obtained. Considering this case in the symmetry
broken phase and shifting the field about the vacuum expectation
value $\bf v \equiv \langle {\bm \phi} \rangle$, one can derive Ward
identities involving $\bf v$. In particular, there is an identity
relating the two and three point 1PI functions (vertices) that can
be compared to the one we derived for broken GI or for broken EGI.
This is (for complete details refer to \cite{ZJ}),
\begin{equation}\label{O(N)}
t^{\alpha}_{li}\Gamma_{ik}^{(2)}(p) +
t^{\alpha}_{ki}\Gamma_{il}^{(2)}(p) +v_j
t^{\alpha}_{ji}\Gamma_{ikl}^{(3)}(p)(0,p,-p),
\end{equation}
where the $t^{\alpha}$, $\alpha= 1,2,...,N$ are $N\times N$ real
antisymmetric matrices: the generators of the $O(N)$ Lie algebra.
These of course have no analogue in our field theory, as the
Galilean transformation is abelian, whereas $O(N)$ is a nonabelian
symmetry group. Apart from these important Lie algebra factors, this
identity holds for the triple vertex $\Gamma^{(3)}$ at zero
four-momentum transfer, an important distinction to Eq.(\ref{WTIc})
which holds for zero 3-momentum transfer but for arbitrary finite
frequency transfer. From the asymmetry of the $t^{\alpha}$ we can
write $t^{\alpha}_{ki} = - t^{\alpha}_{ik}$ and so express the first
two terms in Eq.(\ref{O(N)}) as a \textit{difference}, but there is
no obvious limit to take here that would yield even a formal
derivative, thus making comparison to the zero-momentum and zero
frequency transfer Ward identity in Eq.(\ref{limitWTI}) difficult.
From these considerations, we conclude that the Ward identities
resulting from broken Galilean and broken extended Galilean
invariance bear a closer resemblance to those arising in QED, than
to the those from the above globally invariant N-component scalar
field theory.

\section{\label{sec:vertex}Implications for vertex renormalization}

The vertex function $\Gamma^{(2,1)}$ is associated with the
nonlinear or advective term in the Navier-Stokes equation. In the
action formalism Eq.(\ref{class}), this vertex is given at
tree-level by the nonlinear or advective term multiplied by the
conjugate field: $\sigma_k P_{kj}(\nabla ) \frac{\partial
(V_lV_j)}{\partial x_l}$. It is thus a three-legged object, built up
from one conjugate field and two velocity fields, and this fact is
reflected when writing out the three arguments of
wavevector-frequency pairs. Of course, both wavevector and frequency
are independently conserved at the vertex, so it is in general a
function of two independent wavevector-frequency pairs (see e.g.,
Ref.\cite{Mou} for the elements of Navier-Stokes diagrammatic
perturbation theory).  The most general form of the vertex taking
into account corrections and possible renormalization effects must
reflect this fact. We can separate out the tree level or zero loop
contribution, and thus the term $\Lambda$ represents all the
possible (higher-loop) corrections:
\begin{equation}\label{generalvertex}
\Gamma^{(2,1)}_{ijk}(\mathbf{q},\omega;\mathbf{k},\nu;\mathbf{q-k},\omega-\nu)
= ik_iP_{jk}(\mathbf{k}) + \Lambda_{ijk}(
\mathbf{q},\omega;\mathbf{k},\nu;\mathbf{q-k},\omega-\nu).
\end{equation}
In Eq.(\ref{generalvertex}) our convention is that the middle
argument corresponds to the incoming wavevector and frequency
carried by the conjugate field, whereas the first and third
arguments correspond to the wavevector/frequency pair carried by the
two velocity fields that meet at the vertex.

In a similar fashion, the exact inverse response function can be
written as follows:
\begin{equation}\label{generalresponse}
\Gamma^{(1,1)}_{jk}(\mathbf{k},\omega;-\mathbf{k},-\omega) =
\left[-i\omega + \bar \nu(\omega,k)k^2 + \Sigma(\omega,k)
\right]P_{jk}(\mathbf{k}),
\end{equation}
where $\bar \nu(\omega,k)$ and $\Sigma(\omega,k)$ denote a
renormalized viscosity and a ``mass" term, respectively. At tree
level, this corresponds to the term in the action $\sigma_k
\Big(\frac{\partial V_k}{\partial t} - \nu_0 \nabla^2 V_k\Big)$. So,
the inverse response function $\Gamma^{(1,1)}$, is a two-point
object (built from one conjugate and one velocity field) and thus a
function of just one wavevector and one frequency \cite{Mou}. The
Ward identities derived above in Eqs.(\ref{WTIc},\ref{limitWTI})
impose certain constraints on these three functions $\bar \nu$,
$\Sigma$ and $\Lambda$. In other words, they relate the nonlinear
and linear parts of the Navier-Stokes equation. First consider
implications of extended Galilean invariance. Inserting
Eqs.(\ref{generalvertex},\ref{generalresponse}) into Eq.(\ref{WTIc})
implies that
\begin{eqnarray}\label{conseq1}
k_nP_{jk}(\mathbf{k})&&\left[ (\bar \nu(\omega+\nu,k)-\bar
\nu(\nu,k))k^2 + (\Sigma(\omega+\nu,k) -\Sigma(\nu,k))\right]
\nonumber \\
&=& -\omega \Lambda_{njk}(\mathbf{0},\omega;\mathbf{k},\nu;
-\mathbf{k},\omega - \nu).
\end{eqnarray}
Thus, a vertex correction (i.e., right-hand side of
Eq(\ref{conseq1})) can arise provided that the renormalized
viscosity and/or the mass term are frequency dependent functions,
according to the left-hand side of this relation. However, recall
that the reference system is being subject to arbitrary rectilinear
accelerations, and this noninertial bulk motion does introduce an
explicit time (and hence frequency) dependence into the system.
Nevertheless, the correction to the vertex in this case is the
\textit{specific} one indicated in Eq.(\ref{conseq1}). We see that
the vertex correction that enters into this relation is the one for
which one of the two velocity legs carries \textit{zero} wavevector
$\mathbf{0}$ but finite frequency $\omega$, i.e., this leg
corresponds to a time-dependent velocity zero-mode.

Next, we come to the implication of Galilean invariance. This can be
had by either inserting
Eqs.(\ref{generalvertex},\ref{generalresponse}) into
Eq.(\ref{limitWTI}) or by taking the zero frequency limit $\omega
\rightarrow 0$ directly in Eq.(\ref{conseq1}). Either way, we obtain
\cite{BH}
\begin{equation}\label{conseq2}
k_n \frac{\partial}{\partial \nu}\left[\bar \nu(\nu,k)k^2 +
\Sigma(\nu,k)\right]P_{jk}(\mathbf{k}) =
-\Lambda_{njk}(\mathbf{0},0; \mathbf{k},\nu;-\mathbf{k},-\nu).
\end{equation}
As stated in \cite{BH}, as a model of stationary forced turbulence,
neither the viscosity nor the mass term will depend on frequency, so
that for stationary random forcing, the constraint
Eq.(\ref{conseq2}) implies that
\begin{equation}\label{GIimplicvertex}
\Lambda_{njk}(\mathbf{0},0; \mathbf{k},\nu;-\mathbf{k},-\nu) = 0.
\end{equation}
This specific vertex correction is in fact zero, and it corresponds
to the situation in which one of the two velocity legs carries zero
wavevector $\mathbf{k} = \mathbf{0}$ and zero frequency $\omega =
0$, and this corresponds to a velocity zero mode.

These considerations are important. There has been a longstanding
question about the range of validity of the vertex Ward identity and
the vertex non-renormalization property it implies \cite{FNS,Teo,Toma,Mou,DM,Mc,BH,BH2}.
In \cite{BH2}, we made a definitive resolution of this problem by discovering that
the dynamic functional for Navier-Stokes theory possesses an
underlying and fundamental BRS symmetry. Since the fields of the BRS
symmetry involve only the zero mode, that demonstration makes
explicit that the associated vertex Ward identity is valid only at
exactly zero momentum transfer $\mathbf{k = 0}$, and not in some
small limiting region (i.e., for $k \rightarrow 0$) around this. The
statement that vertex renormalization is \textit{not} constrained by
Galilean invariance was put forward somewhat earlier in \cite{Mc},
using physical arguments and employing the Reynolds decomposition.
Our derivation in \cite{BH2} validates the assertions made in
\cite{Mc} through a mathematically precise and complete construction
(see statements in Ref. \cite{Eyink} concerning the mathematical
rigor of the path integral formalism).  Except for the zero mode,
neither Galilean nor extended Galilean invariance constrain vertex
renormalization.

\section{\label{sec:disc}Discussion}

It has been known for a long time that the randomly forced
Navier-Stokes (NSE) equation can be cast in terms of a path integral
\cite{DeDom,Mou,Teo,Toma,DM,Eyink,BH,AAV}. The underlying symmetries
of this equation can then be treated at the level of the functional
integral. In particular, the Galilean invariance of the NSE is
formally analogous to global gauge invariance in quantum field
theory. The well established methods for gauge fixing can then be
brought to bear and used to restrict the sum over histories of
Navier-Stokes dynamics to pick out one inertial frame. This gauge
fixing was in fact used recently to eliminate an infinite number of
spurious correlation functions that are implied by the non-gauge
fixed functional \cite{BH2}. In addition, the evaluation of
non-Galilean invariant quantities (e.g., $n$-point correlations of
velocity fields) in the standard path integral Eq.(\ref{Z}) leads to
over-counting of configurations and spurious relations, so we must
fix the gauge, that is,  break the Galilean invariance. If in the
future, simulations are carried out using such functionals, our
gauge-fixing procedure could possibly provide some numerical
advantages in allowing for a faster and more efficient code.
However, once the functional has been gauge-fixed, it is crucial to
ascertain whether the well known and oft-cited Ward identity for the
inverse response and vertex functions
\cite{FNS,Teo,Toma,Mou,DM,Mc,BH} remains valid or not. Regarding
this question, it is important to recognize that the gauge-fixed
theory possesses a Becchi-Rouet-Stora (BRS) symmetry \cite{BRS}
which when brought out and exposed, restores a type of hidden
Galilean invariance. The celebrated Ward identity is therefore
re-established, but it is now understood to be a direct consequence
of this BRS invariance, and \textit{not} of the original Galilean
invariance.

We emphasize that our BRS symmetry follows from gauge fixing and
should not be confused with other kinds of BRS invariance, such as
those treated for example, in chapter 16 of Zinn-Justin's book
\cite{ZJ}. In these latter cases, it is the stochastic differential
equation itself that is regarded as a local constraint equation for
the field variable. The determinant of the Jacobian of this
stochastic equation is then expressed as a Grassmann integral over
ghost fields in the path integral representation of the generating
functional for the correlation functions. An effective action is
then identified, and it is this action that is shown to possess a
simple kind of BRS symmetry, and sometimes even a kind of
supersymmetry. This differs distinctively from our treatment here,
where by contrast, the BRS symmetry emerges from an attempt to
correct an infinity in the NSE dynamic functional arising from the
underlying Galilean (and extended Galilean) invariance of the
theory. The ghost and BRS formalism provides the natural language in
which to establish the Ward identities. We are not required to
``compensate"  any degrees of freedom as in quantum gauge theories.
Our use of the ST identities in the stochastic field theory of
randomly stirred flows has an entirely different significance to the
one encountered in quantum field theories.

The NSE is invariant as well under rectilinear but otherwise
arbitrary frame accelerations. This extended Galilean invariance
(EGI) has been considered previously in various differing contexts
\cite{LL,Ivash,Tong,Shah}. Just as for Galilean invariance, EGI can
be regarded as a gauge symmetry at the level of the path integral.
This higher gauge symmetry can be fixed, which corresponds to
choosing one rectilinearly accelerating reference frame. Just as
before, the gauge fixed theory possesses an underlying BRS symmetry
which when invoked, leads to a generalized Ward identity for the
inverse response and vertex functions. This new Ward identity
reduces to the standard one in the limit of zero frame acceleration.
It contains additional information regarding the vertex
renormalization implied by arbitrary rectilinear frame
accelerations, that is not provided by the "standard" Ward identity.
Extended Galilean invariance is of interest in its own right, for by
the principle of equivalence \cite{Weinberg}, it corresponds to the
invariance of the stirred flows under arbitrary time-dependent
unidirectional background gravitational fields.

The standard Ward identity has been appealed to on numerous
occasions to make statements about the nonrenormalization of the
advective or inertial term in the Navier-Stokes equation
\cite{FNS,Teo,Toma,Mou,DM,Mc,BH}. Recently, McComb claimed that
Galilean invariance does not at all constrain the vertex
renormalization \cite{Mc}. To quote his paper verbatim, ``Galilean
invariance has been used as the justification of Ward identities,
which in turn lead to the conclusion that in the perturbative
renormalization group (RG) the vertex is not renormalized".
Nevertheless, it is argued in Ref.[12] that ``vertex renormalization
is not constrained by this (Galilean) symmetry". Section
\ref{sec:vertex} of this paper spells out explicitly what both the
``standard" and the generalized Ward identities imply for the
corrections to the bare, tree-level vertex. We emphasize there how
our results, derived from field theory methods, complement, validate
and confirm the assertions made in Ref [12]. To test the claim put
forward in \cite{Mc}, one must go back to the Ward identity, as well
as its generalized version, and examine carefully what constraints
are actually imposed by them. The field-theoretic formalism employed
here is needed to work out the consequences of breaking both
Galilean invariance and its extension.  The final goal of this is to
understand the vertex renormalization problem. The constraints on
the vertex that follow from Galilean invariance and from extended
Galilean invariance (more precisely, that follow from the respective
BRS symmetries of the gauge-fixed theories) are quite weak. In the
case of Galilean invariance, the vertex correction that ``couples"
to the zero mode of the system is constrained to be zero. For EGI,
the vertex correction that involves the system's bulk acceleration
is computable in terms of the frequency dependent viscosity and mass
terms of the renormalized inverse response function. Galilean
invariance and EGI constrain only the spatial \textit{zero mode} of
the vertex but none of the higher wavenumber modes. This information
is new and impacts directly on the physics of randomly stirred
flows. These points can be appreciated from the diagrammatic
comparison of the corresponding identities from QED and the field
theory of the randomly stirred NSE in Sec \ref{sec:QED}.

It is possible to use the functional Eq(\ref{Z}) without explicitly
fixing the zero-mode for deriving perturbative expansions, as has
been done up to now. Thus by implicitly fixing the zero mode, the
perturbation expansions from the functional are consistent with
those based directly on the NS equation itself. However, in looking
for exact relations between correlation functions from the
functional, unless one is careful, the problem of the zero mode then
shows up and can lead to incorrect results. This mistake precisely
underlies the confusion that has been generated over three decades
in regards vertex renormalization claims.  We have demonstrated
here, following on from \cite{BH2}, how to explicitly fix the zero
mode via the analogy we identified with gauge fixing, and thus
obtain a formally well defined functional.

Although we focussed attention here exclusively on the randomly
forced Navier-Stokes equation, the gauge fixing procedure in this
paper can be applied to other Galilean invariant theories such as
the KPZ equation of random surface growth \cite{KPZ},
magnetohydrodynamics and the stochastic Burgers equation
\cite{BFKL}. In \cite{BFKL}, a saddle point approximation is applied
to the path integral for the Burgers equation to calculate the tails
of the probability density function for the velocity. A special
feature of that problem is a symmetry of gauge-invariance type. In
fact, the Burgers action is seen to be invariant under an
\textit{extended} Galilean transformation. This gauge degree of
freedom is then integrated over by the standard Faddeev-Popov trick.
The aim in \cite{BFKL} was to show that the fluctuations around the
instanton are free from infrared divergences. Once the action for
the Burgers equation is gauge fixed and its EGI broken in this way,
then there should be no spurious correlators there either. We
moreover conjecture that there is a corresponding BRS symmetry that
restores the EGI of the gauge-fixed Burgers dynamic functional with
a subsequent Ward identity following as a consequence of this
symmetry.

A few comments regarding our use of gauge-fixing terminology are in
order. With respect to the terminology as employed in (quantum)
field theory,  a local symmetry is one in which the associated field
transformation involves parameters that depend on the spatial
coordinate x. This is not the case for the Galilean transformation
Eqs. (\ref{galZ11}-\ref{galZ41}), and so it is correct to classify
this as a global (or rigid) transformation. The Galilean
transformation is a global space and time transformation between two
separate inertial frames of reference. In standard field theory
texts (see e.g., \cite{ZJ} and \cite{Ramond}) the local
transformations are also denoted as gauge transformations. Even
though Galilean invariance is a global symmetry, we find it to
useful to employ the terminology of gauge invariance and gauge
fixing in this paper, especially since our breaking of this global
symmetry can be handled with exactly the same methods (FP
determinant, BRS, etc), as employed in gauge theories.  Moreover,
for extended Galilean invariance, the associated transformation is
indeed local in time and so does resemble more closely a gauge
theory. Note also that the convective derivative in the NS equation
is invariant under a Galilean transformation, see e.g., the left
hand side of Eq.(\ref{accelNSE}). This reminds one of the invariance
of the U(1) covariant derivative under a gauge transformation, where
for EGI, the function $\lambda(t)$ plays the role of the arbitrary
gauge function of QED. Since the steps for symmetry breaking in the
path integral very closely resemble the methods employed for gauge
fixing (see Appendix A), and since a BRS invariance results with all
its implications for Ward identities, we feel that it is useful and
instructive to use the terminology of broken gauge invariance in
this paper, albeit perhaps in a slightly loose manner.

In summary this work and \cite{BH2} have uncovered a new symmetry in
the stirred NSE and for related Galilean invariant stochastic
equations. Exploiting this symmetry may be useful in performing
Monte Carlo simulations of the path integrals and actions
corresponding to these systems \cite{HJMU,Doben}. Moreover, given
the deep significance of symmetries in physical problems, other
applications may benefit from recognizing this invariance.

\begin{acknowledgments}
Support was provided to A.B. by the UK Science and Technology
Facilities Council (STFC) and D.H. acknowledges the Grant
AYA2006-15648-C02-02 from the Ministerio de Ciencia e Innovaci\'{o}n
(Spain).
\end{acknowledgments}

\appendix
\section{\label{appendix} Gauge Fixing}

The steps outlined here closely parallel those used for gauge fixing
in quantum field theories \cite{Pokorski} and are adapted to the
case at hand.  We first introduce the functional
$\Delta_f[\mathbf{V}]$ by the following equation:
\begin{equation}\label{FP1}
1 = \Delta_f[\mathbf{V}]\int d\mathbf{c}
\,\delta[\mathbf{f}(\mathbf{V^{c}})],
\end{equation}
and $\mathbf{V^{c}}$ denotes the result of the Galilean
transformation (GT) applied to the velocity field $\mathbf{V}$; see
Eq.(\ref{galZ11}), and the integral is over all constant boost
velocities. We assume that the equation $\mathbf{f}(\mathbf{V^{c}})=
0$ has exactly one solution $\mathbf{c}$ for any initial
configuration $\mathbf{V}$. If we apply a second GT, i.e. we next
transform to a new inertial frame (double-prime) moving with a
velocity $\mathbf{b}$ with respect to the primed frame, then the
transformation rules Eqs.(\ref{galZ11}-\ref{galZ41}) tell us that $
\mathbf{x}' = \mathbf{x}'' + \mathbf{b}t'$ and $t' = t''$ and
\begin{eqnarray}\label{secondGT}
(\mathbf{V^{c}})^{\mathbf{b}}(\mathbf{x},t) &=& ({\mathbf V'}({\bf
x'},t') + \mathbf{c})^{\mathbf{b}}, \nonumber
\\
&=& \mathbf{V}''(\mathbf{x}'', t'') + \mathbf{c + b}, \nonumber
\\
&=& \mathbf{V}''(\mathbf{x}-(\mathbf{c + b})t,t) + \mathbf{c + b} =
(\mathbf{V^{b}})^{\mathbf{c}}(\mathbf{x},t) \equiv \mathbf{V^{c+b}}.
\end{eqnarray}
This exercise is needed to prove that  the Fadeev-Popov (FP)
determinant $\Delta_f[\mathbf{V}]$ is invariant. The invariance of
the FP determinant is needed in turn, to ensure that the volume of
the ``gauge-group" orbit can be factored out from the functional
(see, e.g., \cite{AbersLee}). The first requirement follows from
Eq.(\ref{FP1}) using Eq.(\ref{secondGT}):
\begin{equation}\label{FPinvar}
\Delta_f^{-1}[\mathbf{V^b}] = \int d^3\mathbf{c} \,
\delta^3(\mathbf{f}((\mathbf{V^{b}})^{\mathbf{c}})) = \int
d^3(\mathbf{c + b}) \, \delta^3(\mathbf{f}(\mathbf{V^{c+b}})) =
\Delta_f^{-1}[\mathbf{V}].
\end{equation}
Then, repeating the standard manipulations \cite{Pokorski}, we can
prove that the volume of the gauge group indeed factorizes out to
produce an overall infinite constant. Insert Eq.(\ref{FP1}) into the
functional Eq.(\ref{Z}) apply a Galilean transformation, then using
Eq.(\ref{FPinvar}) we obtain
\begin{equation}\label{pathint1}
\Big(\int d\mathbf{c}\Big)\int [D{\bf V}][D \bm{\sigma}]
\Delta_f[\mathbf{V}]\delta[\mathbf{f}(\mathbf{V})]\exp \{
-S[\mathbf{V}, \bm{\sigma}]\}.
\end{equation}
We still need to actually calculate the FP determinant. To calculate
it, recall \cite{AbersLee} it is sufficient to do so for
infinitesimal ``gauge"-transformations (so, infinitesimal GT's):
\begin{equation}\label{FPdet}
\Delta_f[\mathbf{V}] = \det\frac{\delta f(\mathbf{V^{c}})}{\delta
\mathbf{c}}|_{\mathbf{c} = \mathbf{0}} = \det \Big(\frac{\partial
f_i}{\partial c_j}\Big)|_{\mathbf{c} = \mathbf{0}}.
\end{equation}
We remark that the FP determinant Eq.(\ref{FPdet}) is an ordinary
discrete matrix determinant: it is not a \textit{functional}
determinant. This is because the gauge parameter ($=$ boost
velocity) is an ordinary constant vector, not a field. Next, we
consider the class of ``gauge" conditions of the form
$\mathbf{f}(\mathbf{V}) - \mathbf{U}(\mathbf{x},t) = 0$, where
$\mathbf{U}(\mathbf{x},t)$ is an arbitrary function. The FP
determinant is as before because $\mathbf{U}(\mathbf{x},t)$ is
unaffected by a GT. We make use of this feature to replace the delta
function in Eq.(\ref{pathint1}) by some other function(al) which may
be more convenient for practical calculations. So, in this gauge the
path integral Eq.(\ref{pathint1}) becomes
\begin{equation}\label{pathint2}
\Big(\int d\mathbf{c}\Big)\int [D{\bf V}][D \bm{\sigma}]
\Delta_f[\mathbf{V}]\delta[\mathbf{f}(\mathbf{V}) -
\mathbf{U}(\mathbf{x},t)]\exp \{ -S[{\bf V}, \bm{\sigma}]\}.
\end{equation}
This expression is independent of $\mathbf{U}(\mathbf{x},t)$, so we
can integrate over an arbitrary weight functional. As usual, a
popular choice is the exponential \cite{Pokorski}:
\begin{equation}\label{popchoice}
G[\mathbf{U}] = \exp\Big( -\frac{1}{2\xi}\int d\mathbf{x}\, dt \,
\mathbf{U}^2(\mathbf{x},t)\Big),
\end{equation}
for real parameter $\xi > 0$. For the final step, integrate
Eq.(\ref{pathint2}) over $\mathcal{D}\mathbf{U}$ to obtain
\begin{equation}\label{gfixed}
\int [D{\bf V}][D \bm{\sigma}] \Delta_f[\mathbf{V}]\exp \{ -S[{\bf
V}, \bm{\sigma}] -\frac{1}{2\xi}\int d\mathbf{x}\, dt \,
(\mathbf{f}(\mathbf{V}))^2 \}.
\end{equation}
This is the gauge fixed dynamic functional for randomly stirred
incompressible fluids expressed for the gauge function $\mathbf{f}$.

\section{\label{sec:gammaexp} Expansion for $\Gamma$}

We write out the first few terms of the functional Taylor series for
the effective action in Eq. (\ref{Gamma2}). We display only those
terms actually needed to derive the Ward identity in Eq.
(\ref{WTIk2}) in this paper. In wavevector and frequency space these
are:
\begin{eqnarray}\label{GammaTaylor}
&&\Gamma[\mathbf{V}^{cl},\mathbf{\sigma}^{cl}] = \cdots \int
d\mathbf{q}_1 d\omega_1 \int d\mathbf{q}_2 d\omega_2 \,
V_{\alpha}^{cl}(\mathbf{q}_1,\omega_1)\sigma_{\beta}^{cl}
(\mathbf{q}_2,\omega_2)\Gamma_{\alpha \beta}^{(1,1)}(-\mathbf{q}_1,-\omega_1;-\mathbf{q}_2,-\omega_2)\nonumber \\
&+& \frac{1}{2}\int d\mathbf{q}_1 d\omega_1 \int d\mathbf{q}_2
d\omega_2 \int d\mathbf{q}_3 d\omega_3 \,
V_{\alpha}^{cl}(\mathbf{q}_1,\omega_1) V_{\beta}^{cl}
(\mathbf{q}_2,\omega_2)\sigma_{\gamma}^{cl}
(\mathbf{q}_3,\omega_3)\times
\nonumber \\
&&\Gamma_{\alpha \beta
\gamma}^{(2,1)}(-\mathbf{q}_1,-\omega_1;-\mathbf{q}_2,-\omega_2;
-\mathbf{q}_3,-\omega_3) \nonumber \\
&+& \ldots
\end{eqnarray}
%

\section{\label{sec:box} Space-time box }

The action and dynamic functional can be regularized by enclosing
the system in a spatially and temporally bounded space-time box.
This regularization admits a corresponding discrete version of
Galilean invariance, implying the box Ward identity derived above in
Sec \ref{sec:boxward} from which we can demonstrate the appearance
of spurious relations, in complete parallel to the continuum case
treated in \cite{BH2}. Our box is defined from coordinates $x = 0 -
L$, $y= 0 - L$, and $z= 0 - L$ and time defined from $t=0 - T$,
subject to periodic boundary conditions in space and time.

We expand the fields as
\begin{equation}
v_{\alpha}({\bf x}, t) = \sum_{{\bf n},j} \left(\frac{1}{L}\right)^3
\frac{1}{T} {\tilde v}_{\alpha}({\bf k},\omega) \exp[i \frac{2
\pi}{L}(n_ 1 x_1 + n_2 x_2 + n_3 x_3) - i \frac{2 \pi}{T}j t ],
\end{equation}
\begin{equation}
\sigma_{\alpha}({\bf x}, t)= \sum_{{\bf n}.j}
\left(\frac{1}{L}\right)^3 \frac{1}{T} {\tilde \sigma}_{\alpha}({\bf
k},\omega) \exp[i \frac{2 \pi}{L}(n_ 1 x_1 + n_2 x_2 + n_3 x_3) -i
\frac{2 \pi}{T}jt ],
\end{equation}
and
\begin{equation}
D_{\alpha \beta}({\bf x})= \sum_{{\bf n}} \left(\frac{1}{L}\right)^3
{\tilde D}_{\alpha \beta}({\bf k}) \exp[i \frac{2 \pi}{L}(n_ 1 x_1 +
n_2 x_2 + n_3 x_3)].
\end{equation}
The following integrals over the box are used below:
\begin{eqnarray}\label{boxintegrals}
\int_{0}^{T} dt\, \exp[-i\frac{2\pi}{T} j t] &=& T\delta_{j,0}, \\
\int_{0}^{L} d^3\mathbf{x} \, \exp[i\frac{2\pi}{L}
\mathbf{n}\cdot\mathbf{x}] &=& L^3\delta_{\mathbf{n},\mathbf{0}}.
\end{eqnarray}

In ${\bf k}, \omega$ space we will express the velocity and
auxiliary fields in terms of dimensionless quantities as
\begin{equation}
{\tilde v}_{\alpha}({\bf k}, \omega) = v_{\alpha}({\bf n},j) L^4,
\end{equation}
\begin{equation}
{\tilde \sigma}_{\alpha}({\bf k}, \omega) = \sigma_{\alpha}({\bf
n},j) \frac{T^2}{ L},
\end{equation}
and
\begin{equation}
{\tilde D}_{\alpha \beta}({\bf k}) = D_{\alpha \beta}({\bf n})
\frac{L^5}{T^3}.
\end{equation}

For purposes of dimensional counting, let the spatial dimensions be
expressed as $d(x) = L$ and temporal as $d(t) = T$, which also means
that $d(k) = L^{-1}$ and $d(\omega) = T^{-1}$. The action is
dimensionless $d(S)=0$.  Then from examining individual terms in the
action we find $d({\tilde v}({\bf k}, \omega)) = L^4$, $d({\tilde
\sigma}({\bf k},\omega)) = T^2/L$ and $d(\nu) = L^2/T$. In
configuration space $d(v({\bf x},t)) = L/T$ and $d(\sigma({\bf
x},t))=T/L^4$.

Thus the action Eq.(\ref{class}) in discrete wavenumber and
frequency coordinates is:
\begin{eqnarray}\label{discreteS}
S & = & \frac{1}{2}\sum_{\bf n}\sum_j \sigma_{\alpha}(-{\bf n},-j)
D_{\alpha \beta}({\bf n}) \sigma_{\beta}({\bf n},j)
\nonumber \\
& - & i \sum_{\bf n}\sum_j \sigma_{\alpha}(-{\bf n},-j) [(-2\pi ij +
{\bar \nu} n^2) v_{\alpha}({\bf n},j)
\nonumber \\
& - & 2\pi {\bar M}_{\alpha\beta\gamma}({\bf n}) \sum_{\bf m}\sum_l
v_{\beta}({\bf n}-{\bf m},j-l) v_{\gamma}({\bf m},l)] \ ,
\end{eqnarray}
where we defined the dimensionless viscosity ${\bar \nu}$ as ${\bar
\nu} = \nu_0 (2\pi)^2 T/L^2$ and
\begin{equation}
{\bar M}_{\alpha\beta\gamma}({\bf n}) = \frac{1}{2i} [n_{\beta}{\bar
P}_{\alpha\gamma}({\bf n}) + n_{\gamma}{\bar P}_{\alpha\beta}({\bf
n})] \ ,
\end{equation}
where
\begin{equation}
{\bar P}_{\alpha\beta}({\bf n}) = \delta_{\alpha\beta} -
\frac{n_{\alpha}n_{\beta}}{n^2}.
\end{equation}
In the equation Eq. (\ref{discreteS}), everything is
dimensionless.

The Fourier transform of the Galilean transformation between prime and unprimed frames,
analogous to Eqs.(\ref{galZ11}-\ref{galZ41}), and expressed in discrete
dimensionless coordinates is
\begin{eqnarray}
{\bf n}' & = & {\bf n},
\nonumber \\
j' & = & j - {\bf {\bar c}} \cdot {\bf n},
\nonumber \\
v_{\alpha}({\bf n},j) & = & v'_{\alpha}({\bf n},j - {\bf {\bar c}}
\cdot {\bf n}) +{\bar c}_{\alpha} \delta^3({\bf n})\delta(j),
\nonumber \\
\sigma_{\alpha}({\bf n},j) & = & \sigma'_{\alpha}({\bf n},j - {\bf
{\bar c}} \cdot {\bf n}), \label{kgaldis}
\end{eqnarray}
where we defined the dimensionless boost velocity ${\bf {\bar c}}$
as ${\bf c} = {\bf {\bar c}} L/T$. In general ${\bf {\bar c}} \cdot
{\bf n}$ must be an integer for all integers ${\bf n}$, which thus
means all three coordinates of the discretized boost velocity ${\bf
{\bar c}}$ must be integers.

With the discrete Galilian transformations Eq. (\ref{kgaldis}) in
hand we demonstrate that the discrete action Eq. (\ref{discreteS})
is invariant. Begin with the term
\begin{eqnarray}
& & \sum_{\bf m}\sum_l v_{\beta}({\bf n}-{\bf m},j-l)
v_{\gamma}({\bf m},l) \stackrel{Gal-
Trans}{\Rightarrow} \nonumber \\
&=& \sum_{\bf m}\sum_l v'_{\beta}({\bf n}-{\bf m},j-l -{\bf {\bar
c}} \cdot[\mathbf{n}-\mathbf{m}]) v'_{\gamma}({\bf m},l -{\bf {\bar
c}} \cdot\mathbf{m})\nonumber \\
&+& {\bar c}_{\gamma}v'_{\beta}({\bf n},j -{\bf {\bar c}}
\cdot\mathbf{n}) + {\bar c}_{\beta}v'_{\gamma}({\bf n},j -{\bf {\bar
c}} \cdot\mathbf{n}) + {\bar c}_{\beta}{\bar
c}_{\gamma}\delta(\mathbf{n})\delta(j).
\end{eqnarray}
Therefore, we deduce the Galilean transformation of the following
term in Eq.(\ref{discreteS}):
\begin{eqnarray}\label{Mvv}
& & 2\pi i \sum_{\bf n}\sum_j \sigma_{\alpha}(-{\bf n},-j) {\bar
M}_{\alpha\beta\gamma}({\bf n}) \sum_{\bf m}\sum_l v_{\beta}({\bf
n}-{\bf m},j-l) v_{\gamma}({\bf m},l) \stackrel{Gal-
Trans}{\Rightarrow} \nonumber \\
&=& 2\pi i \sum_{\bf n}\sum_j \sigma'_{\alpha}(-{\bf n},-j + {\bf
{\bar c}} \cdot \mathbf{n}) {\bar M}'_{\alpha\beta\gamma}({\bf
n})\Big(\sum_{\bf m}\sum_l v'_{\beta}({\bf n}-{\bf m},j-l -{\bf
{\bar c}} \cdot[\mathbf{n}-\mathbf{m}]) v'_{\gamma}({\bf m},l -{\bf
{\bar c}} \cdot\mathbf{m})\nonumber \\
&+& {\bar c}_{\gamma}v'_{\beta}({\bf n},j -{\bf {\bar c}}
\cdot\mathbf{n}) + {\bar c}_{\beta}v'_{\gamma}({\bf n},j -{\bf {\bar
c}} \cdot\mathbf{n}) + {\bar c}_{\beta}{\bar
c}_{\gamma}\delta(\mathbf{n})\delta(j) \Big)\nonumber \\
&=& 2\pi i \sum_{\bf n}\sum_{j'} \sigma'_{\alpha}(-{\bf n},-j')
{\bar M}'_{\alpha\beta\gamma}({\bf n})\,\sum_{\bf m}\sum_{l'}
v'_{\beta}({\bf n}-{\bf m},j'-l') v'_{\gamma}({\bf m},l')\nonumber \\
&+& 2\pi \sum_{\bf n}\sum_{j'} \sigma'_{\alpha}(-{\bf
n},-j')[\mathbf{\bar c}\cdot\mathbf{n}] v'_{\alpha}({\bf n},j').
\end{eqnarray}
In arriving at the last equality, we have used fluid
incompressibility $\mathbf{n}\cdot\mathbf{v}= 0$, the fact that
${\bar M}'_{\alpha\beta\gamma}({\bf n})\delta(\mathbf{n}) = 0$ as
well as the change of discrete summation variables $j' =  j - {\bf
{\bar c}} \cdot {\bf n}$ and $l' =  l - {\bf {\bar c}} \cdot {\bf
m}$. Next, consider the Galilean transformation of the ``propagator"
term in Eq.(\ref{discreteS}):
\begin{eqnarray}\label{prop}
& - & i \sum_{\bf n}\sum_j \sigma_{\alpha}(-{\bf n},-j) (-2\pi ij +
{\bar \nu} n^2) v_{\alpha}({\bf n},j) \stackrel{Gal-
Trans}{\Rightarrow}
\nonumber \\
& = & -i \sum_{\bf n}\sum_j \sigma'_{\alpha}(-{\bf n},-j +
\mathbf{\bar c}\cdot\mathbf{n}) (-2\pi ij + {\bar \nu} n^2)[
v'_{\alpha}({\bf n},j - \mathbf{\bar c}\cdot\mathbf{n}) + {\bar
c}_{\alpha}\delta(\mathbf{n})\delta(j)]\nonumber \\
&=& -i \sum_{\bf n}\sum_j \sigma'_{\alpha}(-{\bf n},-j +
\mathbf{\bar c}\cdot\mathbf{n}) (-2\pi ij + {\bar \nu} n^2)
v'_{\alpha}({\bf n},j - \mathbf{\bar c}\cdot\mathbf{n}) \nonumber \\
&=& -i \sum_{\bf n}\sum_{j'}\sigma'_{\alpha}(-{\bf n},-j') (-2\pi
ij' + {\bar \nu} n^2) v'_{\alpha}({\bf n},j') - 2\pi \sum_{\bf
n}\sum_{j'} \sigma'_{\alpha}(-{\bf n},-j')[\mathbf{\bar
c}\cdot\mathbf{n}] v'_{\alpha}({\bf n},j').\nonumber \\
&&
\end{eqnarray}
Lastly, the transformation of the ``noise" term in
Eq.(\ref{discreteS}):
\begin{eqnarray}\label{noise}
& & \frac{1}{2}\sum_{\bf n}\sum_j \sigma_{\alpha}(-{\bf n},-j)
D_{\alpha \beta}({\bf n}) \sigma_{\beta}({\bf n},j) \stackrel{Gal-
Trans}{\Rightarrow}
\nonumber \\
&=& \frac{1}{2}\sum_{\bf n}\sum_j \sigma'_{\alpha}(-{\bf n},-j +
\mathbf{\bar c}\cdot\mathbf{n}) D'_{\alpha \beta}({\bf n})
\sigma'_{\beta}({\bf n},j - \mathbf{\bar c}\cdot\mathbf{n})
\nonumber \\
&=& \frac{1}{2}\sum_{\bf n}\sum_{j' }\sigma'_{\alpha}(-{\bf n},-j')
D'_{\alpha \beta}({\bf n}) \sigma'_{\beta}({\bf n},j').
\end{eqnarray}
Adding up the transformed terms
Eqs.(\ref{Mvv},\ref{prop},\ref{noise}) proves that the discrete
action is indeed invariant under the discrete Galilean
transformation Eq(\ref{kgaldis}).

\end{document}